%% file: main.tex
\documentclass{article}

\usepackage[preprint]{neurips_2026}

\usepackage[utf8]{inputenc}
\usepackage[T1]{fontenc}
\usepackage{hyperref}
\usepackage{url}
\usepackage{booktabs}
\usepackage{amsfonts}
\usepackage{amsmath,amssymb,amsthm}
\usepackage{nicefrac}
\usepackage{microtype}
\usepackage{xcolor}
\usepackage{graphicx}
\usepackage{algorithm}
\usepackage{algorithmic}
\usepackage{cleveref}
\usepackage{subcaption}
\usepackage{multirow}
\usepackage{enumitem}
\usepackage{mathtools}

\usepackage{comment}
\usepackage{tikz}
\usetikzlibrary{arrows.meta, positioning}

\usepackage{makecell}

\graphicspath{{figures/}}

\input{math_commands}

\title{Deep Learning-based Algebraic Reynolds Stress Closures for RANS Simulations of Turbulent Flows}

\author{%
  Daniel Dehtyriov \\
  Mathematical Institute\\
  University of Oxford\\
  Oxford OX2 6GG, United Kingdom  \\
  \texttt{daniel.dehtyriov@maths.ox.ac.uk} \\
  \And
  Jonathan F.~MacArt \\
  Aerospace and Mechanical Engineering \\
  University of Notre Dame \\
  Notre Dame, IN 46556, USA \\
  \texttt{jmacart@nd.edu} \\
  \And
  Justin Sirignano \\
  Mathematical Institute \\
  University of Oxford \\
  Oxford OX2 6GG, United Kingdom \\
  \texttt{justin.sirignano@maths.ox.ac.uk} \\
}

\begin{document}

\maketitle
{\let\thefootnote\relax\footnotetext{Accepted at the 40th Conference on Neural Information Processing Systems (NeurIPS 2026). Code: \url{https://github.com/danieldeh/DARSM}}}

\input{sections/abstract}
\input{sections/introduction}
\input{sections/related_work}
\input{sections/method}
\input{sections/experiments}
\input{sections/discussion}

\begin{ack}
Daniel Dehtyriov’s fellowship is supported by the Schmidt AI in Science Fellowship at the University of Oxford. This work is supported by the U.K. Engineering and Physical Sciences Research Council grant EP/X031640/1 and the U.S. National Science Foundation under Award CBET-22-15472. This research used resources of the Oak Ridge Leadership Computing Facility at the Oak Ridge National Laboratory, which is supported by the Office of Science of the U.S. Department of Energy under Contract No. DE-AC05-00OR22725.
\end{ack}

\bibliographystyle{plainnat}
\bibliography{references,references_cr}

\newpage
\appendix
\crefalias{section}{appendix}\crefalias{subsection}{appendix}
\input{sections/appendix_arxiv}

\end{document}

%% file: math_commands.tex
\newcommand{\vu}{\mathbf{u}}
\newcommand{\vU}{\mathbf{U}}

\newcommand{\vx}{\mathbf{x}}

\newcommand{\vlam}{\boldsymbol{\lambda}}
\newcommand{\vtheta}{\boldsymbol{\theta}}

\newcommand{\G}{\mathcal{G}}
\newcommand{\J}{\mathcal{J}}
\newcommand{\R}{\mathcal{R}}
\newcommand{\Lin}{\mathcal{L}}

\newcommand{\dd}[2]{\frac{\partial #1}{\partial #2}}

\newcommand{\Reals}{\mathbb{R}}

\newcommand{\nut}{\nu_t}
\newcommand{\Reb}{\mathrm{Re}_b}

\DeclareMathOperator{\tr}{tr}

\newcommand{\vv}{\mathbf{v}}

%% file: sections/abstract.tex
\begin{abstract}
Turbulence is ubiquitous in engineering and science, yet direct simulation is prohibitively expensive. The Reynolds-averaged Navier-Stokes (RANS) equations provide computational savings exceeding ten orders of magnitude but introduce unclosed terms requiring modelling. Existing machine-learning (ML) closures suffer distribution shift when trained offline on high-fidelity data, and ML models that bypass the governing equations often have limited capability to generalise. We develop a physics-derived deep learning closure for RANS, the Deep Algebraic Reynolds Stress Model (DARSM), which can be trained on one or two flow cases and generalise across an order of magnitude in Reynolds number, to unseen geometries, and across flow regimes. A neural network maps flow invariants to empirical parameters in an implicit algebraic Reynolds stress equation, derived from the Reynolds stress transport equations under the weak-equilibrium assumption, imposing significant physics-based structure on the ML closure. End-to-end optimisation through the governing PDEs and the coupled implicit closure eliminates distribution shift, but both unrolled and implicit automatic differentiation fail on the stiff coupled solver. We derive adjoint equations that exploit the solver's implicit-explicit structure to enable computationally efficient optimisation. On the canonical square-duct and periodic-hill benchmarks, DARSM reduces average test velocity error relative to the baseline RANS by $2$--$4\times$ across Reynolds number, geometries, and flow regimes, with peak case-level reductions of $12\times$. The model trained on attached, anisotropy-dominated flows (square duct) accurately generalises without retraining to separated flows (periodic hills), a regime change in the underlying turbulence physics. DARSM also outperforms established ML methods: offline training, tensor-basis neural networks, field-inversion machine learning, DeepONets, physics-informed neural networks, and an adjoint-trained source-term closure.
\end{abstract}

%% file: sections/introduction.tex
\section{Introduction}
\label{sec:intro}

Computational fluid dynamics underpins predictive simulation across engineering and science, from aerodynamics and energy generation to atmospheric and biological flows. Direct numerical simulation (DNS) of the governing equations is exact but prohibitively expensive; practical methods solve reduced-order equations for coarse-grained quantities, introducing modelled terms that represent the influence of the fluctuating velocities on the flow. When the Navier-Stokes equations are time-averaged (RANS), the resulting equations can be solved at a cost ten orders of magnitude lower than DNS for typical engineering systems. The tradeoff is that time-averaging introduces an unclosed term, the Reynolds stress tensor, which must be modelled. Standard closures use linear eddy-viscosity assumptions, which substitute a scalar field for the full closure and are inaccurate across many flow regimes \citep{pope2000turbulent}. More complex physics-derived closure models exist, but their fixed empirical coefficients limit improvement across the diverse conditions encountered in practice \citep{duraisamy2019turbulence}. Learning better closures from data is therefore a natural objective.

In scientific settings, training data is inherently scarce because it must be generated from experiments or high-fidelity simulations, both of which are expensive and limited in coverage. However, unlike many other applications of machine learning (ML), in scientific fields the governing equations provide critical mathematical structure for learning. Many existing machine learning approaches to turbulence closure do not fully leverage this structure. Methods that bypass the governing equations and learn the flow field directly forfeit the physics structure that makes generalisation possible from limited data. Although these methods can be successful given sufficiently large datasets, their out-of-sample accuracy can be limited when trained on small or medium-size datasets. Methods that retain the governing equations but train the closure offline on DNS data (a common approach) face a different problem. The network is trained on inputs drawn from the reference DNS data, but at deployment (during the RANS simulation) it receives inputs from the solver's own approximate RANS solution, a different region of input space. This mismatch, a \emph{distribution shift} in ML terms, either destabilises the solver or significantly degrades accuracy~\citep{sirignano2020dpm}, as we demonstrate empirically in \Cref{sec:comparison}. What both families lack is a closure form that constrains the network's role and a training procedure that keeps its outputs consistent with the governing equations.

We develop a deep learning anisotropic closure for RANS, the Deep Algebraic Reynolds Stress Model (DARSM), in which neural networks model empirical parameters of an implicit algebraic closure equation derived from the Reynolds stress transport equations (RSTE) under the weak-equilibrium assumption \citep{Rodi1976} (\Cref{sec:closure}; a full derivation is in Appendix~\ref{app:earsm_derivation}). The algebraic equation fixes the tensorial structure of the closure from physics, maximally leveraging the physics available on the coarse RANS mesh and confining the neural networks to a low-dimensional correction of terms that cannot be determined from first principles. To address distribution shift, we train DARSM by optimizing over the entire PDE system of the RANS momentum equations, continuity equation, and coupled implicit algebraic closure equations (which include embedded neural networks to model empirical parameters). Standard automatic differentiation (AD) fails on this stiff PDE system, so we derive a system of implicit adjoint equations designed to leverage the forward numerical solver's implicit-explicit splitting to provide computationally efficient optimization for training DARSM. In particular, fast numerical solvers can be used to solve the implicit adjoint equations due to their special structure.

In contrast to previous adjoint-trained machine learning closures for RANS, DARSM is an anisotropic closure that predicts the full Reynolds-stress tensor rather than a scalar eddy viscosity, which is required wherever the mean-flow dynamics depend on the directional structure of the Reynolds stresses (a regime that spans many engineering and geophysical flows, including those with secondary motions, separation, swirl, or strong streamline curvature). To motivate this, consider the most direct ML approach to anisotropic closure. By the Cayley-Hamilton theorem, any frame-invariant closure can be written as a sum of ten tensor basis elements with scalar coefficients depending on at most five invariants of the mean flow. The tensor-basis neural network (TBNN) of \citet{ling2016reynolds} learns these ten Pope tensor-basis coefficients directly as neural networks of the invariants, guaranteeing frame-invariance but leaving the network to infer the coefficient structure entirely from data. Algebraic Reynolds Stress Models (ARSM) instead derive the coefficient structure from the physics. Under the weak-equilibrium approximation, which assumes that the anisotropic structure of the Reynolds stress equilibrates with the local mean strain on a timescale much shorter than the mean flow varies along streamlines, the RSTE reduce to a coupled nonlinear algebraic equation for the stresses as functions of local mean-flow invariants. The algebraic equation contains empirical parameters that must be modelled. Whereas TBNN must learn nearly all of the physics from the data, DARSM provides a physics-based mapping from these empirical parameters to the tensor-basis coefficients, and this structure enables generalisation from scarce data.

The cost is that the neural network weights are embedded within a set of stiff coupled semi-implicit PDEs, and the loss gradient requires differentiating through the fully converged solver. The iterative PDE solver can itself be viewed as a deep computational graph, and its adjoint is the natural analogue of reverse-mode automatic differentiation. The key difference is that several intermediate numerical steps require solving implicit PDEs (pressure Poisson equation and multiple alternating-direction implicit sweeps for the momentum equation). Direct automatic differentiation, whether through unrolled solver iterations or off-the-shelf implicit differentiation, fails at the grid sizes required for practical turbulence predictions. We derive discrete adjoint equations that exploit the solver's implicit-explicit splitting, with implicit adjoint equations for the forward implicit sub-steps (e.g., the adjoint Poisson equation) and reverse-mode automatic differentiation for the explicit assembly (including all neural-network components). See \Cref{sec:method} for a detailed description of our adjoint method. A key advantage of this approach is that our fast numerical solvers for the forward implicit PDE sub-steps can be re-used for the corresponding adjoint PDE steps in backpropagation. This gives exact gradients at constant memory, independent of the number of solver iterations. This construction extends to any semi-implicit PDE solver.

We evaluate DARSM on two canonical turbulence benchmarks chosen for their structurally distinct physics: flow through a square duct (where the important physics is a weak cross-sectional circulation that standard closures cannot represent at all) and flow over periodic hills (where the flow separates and reattaches in ways standard closures consistently mispredict). DARSM, trained on one or two cases, generalises across an order of magnitude in Reynolds number, across duct aspect ratios, and across periodic-hill geometries, reducing the mean out-of-sample velocity error against the baseline RANS by $2\times$, $3\times$ and $4\times$ respectively, with peak case-level reductions of $12\times$. The duct-trained model also transfers without retraining to the periodic hills, with accuracy on par with a closure trained directly on the hills. The dominant turbulence physics changes from corner-driven secondary motion to separated shear-layer dynamics, a regime change that is not simply a parameter or geometry shift. Of the six alternative ML methods we benchmark, none improves on the baseline closure in mean out-of-sample error on both flows, while DARSM does (\Cref{sec:comparison,sec:hills}).

\paragraph{Contributions:}
\begin{enumerate}[nosep]
\item \textbf{Physics-derived implicit neural closure for turbulence.} We develop and train a new machine learning-based anisotropic closure for RANS. The deep algebraic Reynolds stress model (DARSM) closure is derived from the Reynolds stress transport equations (RSTE), where the neural network models empirical parameters within an approximation for the RSTE which is implicitly coupled with RANS. Nearly all of the model is guided by physics except for the lowest-level Reynolds stress dynamics, allowing the model to better generalize to new flow regimes and geometries. To our knowledge, DARSM is the first physics-derived deep learning algebraic Reynolds stress model and the first adjoint-optimised closure that goes beyond scalar isotropic corrections. \\

\item \textbf{Generalisation from minimal data.} DARSM trained on one or two DNS cases generalises across an order of magnitude in Reynolds number, across duct aspect ratios, and across periodic-hill geometries, reducing the mean out-of-sample velocity error against the baseline RANS by $2\times$, $3\times$ and $4\times$ respectively, with peak case-level reductions of $12\times$. When trained only on square-duct data, DARSM generalises without retraining to periodic hills, a flow regime with separation, reattachment, and anisotropy mechanisms absent from the training data, reducing error $4.6\times$ over the classical baseline (\Cref{sec:duct,sec:hills}). \\

\item \textbf{Discrete adjoint for stiff implicit PDE solvers with embedded neural components.} We derive a hybrid adjoint-automatic differentiation method that reuses the forward solver's own implicit PDE solvers for the implicit adjoint equations and reverse-mode AD for the explicit PDE sub-steps, giving exact gradients at constant memory with no additional solver infrastructure. Unrolled AD and full-autograd adjoints both fail at the grid sizes required. The method extends to other semi-implicit PDE solvers (see \Cref{sec:hybrid} for formulation and scaling, and Appendix \ref{app:cdr} for a second PDE class). \\

\item \textbf{Comparison with existing scientific machine learning methods.}

Five established machine learning methods are evaluated on the same data as benchmarks: tensor-basis neural networks (TBNN) \citep{ling2016reynolds}, the same DARSM architecture but trained offline rather than via adjoint optimization (a-priori NN-EARSM), field inversion machine learning (FIML)~\citep{parish2016paradigm}, DeepONet~\citep{lu2021learning}, and physics-informed neural networks (PINNs)~\citep{raissi2019physics}. We additionally compare against a closure that represents the unclosed Reynolds stress by a neural-network source term in the momentum equations~\citep{kochkov2021machine,sirignano2020dpm}, trained through the solver with the same adjoint as DARSM. DARSM is the only method that improves on the baseline RANS in mean out-of-sample error on both flows, and the comparison shows that neither the physics-derived parameterisation nor solver-in-the-loop training suffices on its own (\Cref{sec:experiments}).

\end{enumerate}

%% file: sections/related_work.tex
\section{Related Work}
\label{sec:related}

\paragraph{Machine learning for turbulence closure.}

The dominant approach in the existing literature on machine learning for turbulence has been \emph{a-priori} training, where closure model parameters are fitted offline to high fidelity reference data \citep{Brunton2020,duraisamy2019turbulence,Sanderse24}.

Given time-averaged DNS data $\overline{u^*}$, the exact values $y^*=y(\overline{u^*})$ of the unclosed terms in RANS can be calculated. Then \emph{a priori} training minimises the supervised loss $\mathcal{L}_{\text{offline}}(\vtheta) =
\| f_{\vtheta}(\phi(\overline{u^*})) - y^* \|^2$, where $f_{\vtheta}$ is the closure network and $\phi$ maps the time-averaged flow state to the network's input features. Once optimized offline, the trained parameters $\vtheta^*$ are frozen. $f_{\vtheta^*}$ is then inserted as a model for the unclosed terms in RANS and the RANS system is numerically solved to obtain a solution $U(\vtheta^*)$. In practice, the RANS solution $U(\vtheta^*)$ is almost always significantly different from the DNS solution, leading to the closure model being evaluated on a different region of the input space than it was trained on (since it was trained on DNS data). That is, while the closure model is trained offline on $\overline{u^*}$, during predictive simulations it instead receives the RANS input $U$. This distribution shift causes the \emph{a priori}-trained closure model to be inaccurate and therefore can substantially limit the accuracy of the RANS predictive simulations \citep{sirignano2020dpm, um2020solver, duraisamy2019turbulence}.

The TBNN model~\citep{ling2016reynolds} uses \emph{a priori} training to fit a tensor basis neural network to DNS, enforcing frame-invariance by construction but suffering distribution shift during predictive simulations. FIML~\citep{parish2016paradigm,singh2017machine} is a two-step procedure in which the adjoint first fits a spatial correction field to a closure coefficient through the solver and a neural network is then trained offline to predict these fields from local flow features. The first step is solver in-the-loop and avoids distribution shift, but the second reintroduces it. Physics-informed neural networks (PINNs)~\citep{raissi2019physics} replace the PDE solve with a neural network trained to minimise the residual of the governing equations and the mismatch with the data. They are flexible for inverse problems and data assimilation, but typically require per-case retraining, and highly accurate solutions for stiff equations remain an open research problem~\citep{krishnapriyan2021,wang2022,ji2021}. The Optimizing a Discrete Loss (ODIL) framework~\citep{Karnakov2023,Karnakov2024} minimises the PDE residual of a discretised function (i.e., a tensor representing the underlying physical system at a discrete set of grid points) plus an error term measuring how well the discretised function matches observed data at a (potentially sparse) set of points. ODIL provides a highly flexible framework for interpolating observed data across the spatial domain while constraining the interpolated function to satisfy the physical laws represented by the PDE and can also leverage Newton-type convergence of classical solvers. Neural operator networks~\citep{lu2021learning,li2021fourier} do not solve the governing PDEs when making predictions, but instead aim to directly learn a map between the physical parameters characterizing a problem and the corresponding solution field. They offer fast amortised inference of the full solution map but need large training sets to generalise. Symbolic closure methods \citep{Weatheritt2017,Schmelzer20,fang2023toward} search for interpretable algebraic expressions via evolutionary or sparse regression, but \emph{a-priori} variants face the same distribution shift, and gradient-free a-posteriori search scales poorly with parameter count. To our knowledge, DARSM is the first physics-derived deep learning algebraic Reynolds-stress model.

\paragraph{End-to-end training over PDE solvers.} DARSM involves a stiff pressure-coupled semi-implicit PDE solver with embedded neural network components, a regime where unrolled AD and off-the-shelf implicit differentiation either fail or scale poorly (\Cref{sec:scaling}). Unrolled discrete adjoints via reverse-mode AD~\citep{um2020solver,List2022,Frezat22,shankar2023differentiable} backpropagate through the solver trajectory; the forward graph is retained so memory scales as $\mathcal{O}(KN)$ in the rollout length $K$ and number of grid points $N$. General-purpose differentiable simulation frameworks~\citep{holl2020phiflow,macklin2022warp} expose solvers to automatic differentiation via this unrolling strategy and inherit the same memory cost. Neural differential equations~\citep{kochkov2021machine,kochkov2024neural} are trained in the same way, with the neural network embedded in the discretised governing equations as a learned correction, as interpolation weights or additive tendencies acting directly on the resolved fields; they have been applied to coarse-grained unsteady flow and to global atmospheric models. The DPM framework~\citep{sirignano2020dpm} instead derives a backward in time adjoint for the time-discretised scheme so that the graph can be discarded at each timestep. The framework introduces this as a general strategy for augmenting a known PDE with a neural-network correction; \citet{kakka2025neural} apply it to RANS, augmenting the scalar eddy viscosity for compressible turbulent jet flames. Continuous adjoints derive the adjoint PDE analytically and discretise it separately, introducing discretisation-consistency error \citep{giles2000introduction}. \citet{stromer2021end} used a continuous adjoint to train a tensor-basis network through the steady RANS equations, learning a linear eddy viscosity from DNS data, and \citet{zhang2022ensemble} trained the same closure with a derivative-free ensemble Kalman method. Steady-state discrete adjoints eliminate the time horizon entirely, with $\mathcal{O}(N)$ memory and discrete exactness, and are the natural analogue of implicit differentiation of equilibrium models~\citep{bai2019deq}. \citet{holland2019fiml} first applied this to neural closure training, correcting a scalar isotropic quantity from experimental measurements on an airfoil within the SU2 framework~\citep{albring2015su2}. \citet{sirignano2023pde} and \citet{Reidl2025} provided convergence theory for PDE-constrained optimisation with neural network terms. \citet{agrawal2024probabilistic} cast closure training as Bayesian inference through a differentiable RANS solver, learning a tensor-basis neural network alongside a stochastic model-discrepancy field. Multi-agent reinforcement learning has been used to train LES closures end-to-end against solver output~\citep{Novati2021, Bae2022}. Adjoint methods for CFD originate in aerodynamic shape optimisation~\citep{jameson1988aerodynamic,giles2000introduction}, where hand-derived adjoints provide gradients with respect to geometric design variables, and the turbulence model is frequently frozen in the adjoint.

We derive a hybrid adjoint-AD method that formulates an implicit adjoint equation for each implicit forward sub-step, solves these implicit adjoint equations with fast numerical solvers that leverage the special structure of the implicit equations, applies AD to the explicit RHS terms and embedded neural network evaluations (allowing for highly flexible neural network architectures), and connects the implicit solves with the explicit AD calculations using additional adjoint equations. A key advantage is that the implicit adjoint equations have the same structure as the corresponding forward implicit sub-steps in the solver. This allows us to re-use the same highly efficient numerical solvers used for the implicit forward equations to numerically solve the implicit adjoint equations. For example, the adjoint equation for the pressure Poisson equation will also be a Poisson equation. The hybrid adjoint-AD method differentiates the fully coupled system, including the turbulence closure, and gives exact gradients at $\mathcal{O}(N)$ memory. Unrolled and off-the-shelf implicit differentiation both fail for the flow cases considered in this paper (see \Cref{sec:hybrid,sec:scaling}).

%% file: sections/method.tex
\section{Method}
\label{sec:method}

\subsection{RANS closure problem}
\label{sec:closure}

The incompressible Navier-Stokes equations govern the motion of fluid flow:
\begin{equation}\label{eq:incompresible}
\partial_t u_i + u_j \partial_j u_i = -\tfrac{1}{\rho}\partial_i p + \nu \partial_j \partial_j u_i, \qquad \partial_i u_i = 0.
\end{equation}
DNS solves these equations exactly, resolving all scales of turbulent motion, but the computational cost is prohibitive. In engineering applications, the quantities of interest are typically time-averaged flow fields. Decomposing each quantity into $u_i = U_i + u'_i$, where $U_i$ is the time-averaged mean and $u_i'$ is the fluctuation around this mean, and then time-averaging yields the Reynolds-averaged Navier-Stokes (RANS) equations for the mean velocity $U_i$ and pressure $P$:

\begin{equation}
    U_j \partial_j U_i
    = -\tfrac{1}{\rho}\partial_i P
    + \partial_j\!\left[\nu(U_{i,j}+U_{j,i}) - \langle u'_i u'_j\rangle\right],
    \qquad \partial_i U_i = 0,
    \label{eq:rans_main}
\end{equation}
where $U_{i,j}\equiv\partial U_i/\partial x_j$. The Reynolds-stress tensor $\langle u'_i u'_j\rangle$ is unknown and must be modelled (the closure problem). The standard Boussinesq closure $\langle u'_i u'_j \rangle \approx -\nu_t (U_{i,j} + U_{j,i}) + \frac{2}{3}k\delta_{ij}$ is an isotropic eddy-viscosity model, in which a single scalar $\nu_t$ aligns the deviatoric Reynolds stress with the mean strain rate. This is structurally inadequate for flows where anisotropy drives the physics, since secondary motions in square ducts are identically zero under the Boussinesq hypothesis, stress-strain misalignment is excluded by construction, and separated flows are poorly predicted~\citep{pope2000turbulent}. 

An algebraic Reynolds stress model (ARSM) is derived from the Reynolds stress transport equation under the weak-equilibrium assumption, which holds when the anisotropy equilibrates with the local strain on a timescale short relative to mean-flow variation along streamlines (Appendix~\ref{app:earsm_derivation}). It retains the anisotropy of the full-scale flow through an implicit nonlinear equation coupling the stress with the mean strain and rotation. The explicit ARSM (EARSM) of \citet{wallin2000explicit} provides the anisotropic correction:
\begin{equation}
    \langle u'_i u'_j\rangle
    = -\nu_t (U_{i,j}+U_{j,i})
    + \tfrac{2}{3} k\, \delta_{ij}
    + k\, a^{(\mathrm{ex})}_{ij}(\mathbf{S}^*, \boldsymbol{\Omega}^*; c_1, c_2).
    \label{eq:stress_main}
\end{equation}
The turbulence scales $k$ (kinetic energy) and $\omega$ (specific dissipation rate) satisfy coupled transport equations~\citep{wilcox1988reassessment}:
\begin{equation}
    U_j\partial_j k = P_k - \beta^* k\omega + \partial_j[(\nu{+}\sigma_k\nut)\partial_j k], \qquad
    U_j\partial_j \omega = \gamma\tfrac{\omega}{k}P_k - \beta_0\omega^2 + \partial_j[(\nu{+}\sigma_\omega\nut)\partial_j \omega],
    \label{eq:komega}
\end{equation}
where $P_k = 2\nut S_{ij}S_{ij} - k\,a^{(\mathrm{ex})}_{ij}S_{ij}$ and $S_{ij} = \tfrac{1}{2}(U_{i,j}+U_{j,i})$ is the mean strain rate.
The EARSM anisotropy $a^{(\mathrm{ex})}_{ij}$ is a nonlinear algebraic function of the normalised strain $\mathbf{S}^* = \frac{\tau}{2}(U_{i,j}{+}U_{j,i})$ and rotation $\boldsymbol{\Omega}^* = \frac{\tau}{2}(U_{i,j}{-}U_{j,i})$ tensors ($\tau = 1/(\beta^*\omega)$), depending on invariants of $(\mathbf{S}^*, \boldsymbol{\Omega}^*)$ and on the pressure-strain parameters $(c_1, c_2)$; the explicit algebraic form is given by \Cref{eq:aex_split}.
The full system has five coupled fields ($U_i, k, \omega$) and seven empirical coefficients $(\beta^*, \beta_0, \gamma, \sigma_k, \sigma_\omega, c_1, c_2)$; the complete derivation is in Appendix~\ref{app:earsm_derivation}.

The stiffness of the coupled $k$--$\omega$ source terms requires semi-implicit treatment.
The system is discretised on a staggered grid and advanced by an ADI scheme in which each pseudo-time step solves a sequence of tridiagonal systems (one per field per direction) followed by a pressure projection, defining the implicit operator $\mathbf{H}$ and explicit right-hand side $\mathbf{g}$ exploited by the adjoint in \Cref{sec:hybrid}; solver details are in Appendix~\ref{app:implementation}.

A gated neural network (architecture in Appendix~\ref{app:nn_arch}) maps local invariants of $(\mathbf{S}^*, \boldsymbol{\Omega}^*)$~\citep{pope1975more,ling2016reynolds} and a turbulent Reynolds number to spatially varying corrections on these seven coefficients. The Deep Algebraic Reynolds Stress Model (DARSM) is the coupled system formed by the RANS equations (\Cref{eq:rans_main}), the implicit ARSM closure (\Cref{eq:arsm_general} in Appendix~\ref{app:earsm}), and the neural network supplying the ARSM's empirical parameters. 

\subsection{Physics-derived Implicit Closure to RANS}
\label{sec:rationale}

An anisotropic neural closure for RANS faces a basic design choice. By the Cayley--Hamilton theorem, the anisotropy decomposes as $a_{ij} = \sum_{n=1}^{10}\beta^{(n)}\hat{T}^{(n)}_{ij}$, with the basis tensors $\hat{T}^{(n)}$ built from $\mathbf{S}^*, \boldsymbol{\Omega}^*$ and the scalar coefficients $\beta^{(n)}$ depending on at most their five invariants (Appendix~\ref{app:earsm}). The tensor-basis neural network of \citet{ling2016reynolds} learns the $\beta^{(n)}$ directly as neural networks of the invariants, which guarantees frame-invariance but leaves the coefficients $\beta^{(n)}$ to be inferred entirely from data. In DARSM, the coefficients $\beta^{(n)}$ are instead given by the ARSM equation, which is derived using physics-based arguments, and only the empirical parameters, $(c_1, c_2)$ within the ARSM and the five coefficients of the $k$--$\omega$ equations, are modelled as neural networks. Solving the ARSM shows that the $\beta^{(n)}$ are rational functions of the invariants and the empirical constants $(c_1, c_2)$. The neural network therefore models only the unknown empirical parameters and leaves the structure of the closure to the known physics of the RANS framework.

This choice changes how network corrections propagate through the closure. Under the linear eddy-viscosity (Boussinesq) hypothesis, $\beta^{(1)} = -2C_\mu$ and $\beta^{(n)} = 0$ for $n \geq 2$, where $C_\mu\approx0.09$ is the eddy-viscosity model coefficient (equivalent to $\beta^*$ in the $k$--$\omega$ convention used here); the sensitivity $\partial\beta^{(1)}/\partial C_\mu = -2$ is independent of the local flow state. Nonlinear eddy-viscosity models~\citep{pope1975more, ling2016reynolds} extend the basis to $\hat{T}^{(2)}$--$\hat{T}^{(4)}$ but retain this property. Their coefficients are $\beta^{(n)} = c_n C_\mu$ with fixed $c_n$, so perturbing any model constant rescales the closure output uniformly across invariant space. The ARSM is structurally different. The rational form of $\beta^{(n)}$ in the invariants and the closure constants makes the sensitivities $\partial\beta^{(n)}/\partial c_{1,2}$ nontrivial functions of the invariants by construction. A neural network perturbing the ARSM's empirical constants therefore operates in coordinates where the physics already dictates how corrections propagate, rather than leaving that structure to be learned.

The principle extends beyond turbulence. For any PDE with an algebraic, physics-derived closure form, embedding the neural network as a perturbation of the closure's empirical constants couples the network's inputs to its corrections through the governing equations, supplying inductive bias that a tensor-basis or direct-field network cannot.

\subsection{Adjoint-optimised training through the solver}
\label{sec:training}

Training the closure requires differentiating a scalar loss through the converged RANS solution with respect to the neural network parameters. Let $\vU \in \Reals^n$ be the full state vector and $\vtheta$ the neural network weights. Given high-fidelity target data $\overline{\vu^*}$ (the time-averaged DNS reference projected onto the RANS grid), we seek $\vtheta$ minimising a loss $\J(\vtheta) = \ell(\vU(\vtheta), \overline{\vu^*})$, where $\vU(\vtheta)$ is the converged solution of the discretised RANS PDEs. The target data $\overline{\vu^*}$ comes from direct numerical simulations (DNS) of the incompressible Navier--Stokes equations, resolved down to the Kolmogorov scale~\citep{pirozzoli2018turbulence,vinuesa2018turbulent,breuer2009flow}. The dependence of $\vU$ on $\vtheta$ is implicit, since changing $\vtheta$ changes the PDE and hence its steady-state solution. Because training and deployment use the same solver, the network is always queried on states the solver produces, which avoids distribution shift.

\paragraph{Gradients via the discrete adjoint.}
The iterative solver solves for a fixed point of
\begin{equation}
    \vU^{(k+1)} = \G(\vU^{(k)}; \vtheta).
    \label{eq:fixedpoint}
\end{equation}
At convergence the residual $\R(\vU; \vtheta) \coloneqq (\G(\vU;\vtheta) - \vU)/\Delta t$ vanishes.
Stationarity of the Lagrangian $\J - \vlam^\top\R$ with respect to $\vU$ gives the discrete adjoint system for $\vlam \in \Reals^n$ (Appendix~\ref{app:lagrangian}),
\begin{equation}
    \left(\dd{\R}{\vU}\right)^\top \vlam = \left(\dd{\J}{\vU}\right)^\top,
    \label{eq:adjoint_system}
\end{equation}
from which the parameter gradient follows as $\nabla_{\vtheta} \J = -\vlam^\top (\partial \R / \partial \vtheta)$.
This is the system given by the implicit function theorem (IFT)~\citep{bai2019deq}, but without requiring $\partial\R/\partial\vU$ to be invertible, which the non-unique pressure prevents. We nevertheless keep the IFT label throughout to align with the machine-learning literature.
Memory is $\mathcal{O}(N)$, independent of the number of forward iterations $K$; unrolled backpropagation through $K$ steps requires $\mathcal{O}(KN)$ memory, and on this stiff system ($K \sim 10^4$) saturates memory limits (\Cref{tab:scaling}).

Solving \eqref{eq:adjoint_system} with an iterative method needs the matrix-vector product $\vv \mapsto (\partial\R/\partial\vU)^\top\vv$ at each step; the matrix itself is never assembled. Reverse-mode automatic differentiation of the forward solver can produce this product, but inaccurately and expensively (\Cref{tab:scaling}).

\paragraph{Hybrid discrete adjoint.}\label{sec:hybrid}

Computing this product on a stiff coupled PDE solver is challenging, because it must be exact and tractable at production grid sizes, and naive AD fails on both counts. Unrolled AD saturates the memory budget at $\mathcal{O}(KN)$, and off-the-shelf implicit differentiation is poorly conditioned. We instead exploit the implicit-explicit structure of any semi-implicit PDE solver. The forward residual is a sequence of $M$ implicit sub-steps (ADI sweeps, the pressure projection, and the coupled $k$--$\omega$ block). We derive the adjoint PDEs for each, which turn into tridiagonal, $2{\times}2$ block-tridiagonal, and symmetric-Laplacian systems with the same structure as the forward systems, so the forward solvers can be reused directly in transpose form to solve the corresponding adjoint equations, whereas generic AD would differentiate each implicit solve through its inner iterations. The explicit assembly of each sub-step, including the embedded network, is differentiated by reverse-mode AD.

\begin{figure}[t]
\centering
\begin{tikzpicture}[
    >={Stealth[length=2mm]},
    box/.style={
        draw, rectangle, rounded corners=2pt,
        align=center, font=\small,
        inner xsep=2mm, inner ysep=2mm,
        line width=0.5pt,
        minimum height=14mm,
        minimum width=42mm
    },
    widebox/.style={
        draw, rectangle, rounded corners=2pt,
        font=\small,
        inner xsep=3mm, inner ysep=2.5mm,
        line width=0.5pt,
        text width=110mm,
        align=flush left
    },
    arrow/.style={->, line width=0.7pt}
]

\node[box, fill=blue!8] (fwd) at (0,0)
    {Forward solve\\
     {\footnotesize $\R(\vU,\vtheta)=0$}};
\node[box, fill=orange!12, right=8mm of fwd] (gmres)
    {Outer GMRES on adjoint\\
     {\footnotesize $(\partial\R/\partial\vU)^\top\vlam = (\partial\J/\partial\vU)^\top$}\\
     {\footnotesize $\;\to\;\vlam$}};
\node[box, fill=green!10, right=8mm of gmres] (extract)
    {Extract gradient\\
     {\footnotesize $\nabla_{\vtheta}\J = -\vlam^\top\,(\partial\R/\partial\vtheta)$}};

\draw[arrow] (fwd) -- (gmres);
\draw[arrow] (gmres) -- (extract);

\node[widebox, fill=blue!8, below=10mm of gmres] (forward)
    {\textbf{Forward (one outer iteration):}
     \;$\vU \;\xrightarrow{\;\G_1\;}\;\vU_1\;\xrightarrow{\;\G_2\;}\;\cdots\;\xrightarrow{\;\G_M\;}\;\G(\vU,\vtheta)$\\
     each $\G_m$: assemble $(\mathbf{H}_m,\mathbf{g}_m)$ using $f_{\vtheta}$;
     solve $\mathbf{H}_m\boldsymbol{\delta}_m = \mathbf{g}_m$; \;$\vU_m = \vU_{m-1} + \boldsymbol{\delta}_m$.\\
     Iterate to $\R = 0$; cache $\{\mathbf{H}_m,\mathbf{g}_m,\boldsymbol{\delta}_m\}$.};

\node[widebox, fill=orange!12, below=6mm of forward] (chain)
    {\textbf{Backward (the chain):}
     $\hat{\vU}_{M}\;\to\;\hat{\vU}_{M-1}\;\to\;\cdots\;\to\;\hat{\vU}_{0}$
     \;\;{\footnotesize ($m$-th arrow: one sub-step, below)}\\[0.5mm]
     \quad $\bullet$ GMRES iter: $\hat{\vU}_M = \vv$; \;return $(\hat{\vU}_0 - \vv)/\Delta t = (\partial\R/\partial\vU)^\top\vv$.\\
     \quad $\bullet$ Gradient pass: $\hat{\vU}_M = \vlam$; \;sum per sub-step $\hat{\vtheta}_m$ (step 3 below).};

\node[widebox, fill=orange!12, below=6mm of chain] (subst)
    {\textbf{One backward sub-step} ($m$-th sub-step adjoint, hybrid). Input $\hat{\vU}_m$; cached $\mathbf{H}_m,\mathbf{g}_m,\boldsymbol{\delta}_m$.\\
     (1) \emph{Implicit (transpose solver)}: \;$\mathbf{H}_m^\top \hat{\mathbf{g}}_m = \hat{\vU}_m$\\
     (2) \emph{Matrix adjoint}: \;$\hat{\mathbf{H}}_m = -\hat{\mathbf{g}}_m\,\boldsymbol{\delta}_m^\top$\\
     (3) \emph{Explicit (AD pass)}: \;$\Psi_m = \hat{\mathbf{g}}_m^\top\mathbf{g}_m + \hat{\mathbf{H}}_m{:}\mathbf{H}_m$, \;AD w.r.t.\ $(\vU,\vtheta)$:
     \;\;$\partial\Psi_m/\partial\vU \to \hat{\vU}_{m-1}$, \;\;$\partial\Psi_m/\partial\vtheta \to \hat{\vtheta}_m$.};

\node[widebox, fill=green!10, below=6mm of subst] (gradbox)
    {\textbf{Gradient extraction:} apply the chain with $\hat{\vU}_M = \vlam$ and sum the sub-step contributions:
     \vspace{-0.5mm}
     \begin{center}
     $\nabla_{\vtheta}\J \;=\; -\vlam^\top\,\dfrac{\partial\R}{\partial\vtheta}
     \;=\; -\dfrac{1}{\Delta t}\,\vlam^\top\,\dfrac{\partial\G}{\partial\vtheta}
     \;=\; -\dfrac{1}{\Delta t}\sum_{m=1}^{M} \hat{\vtheta}_m.$
     \end{center}};

\end{tikzpicture}
\caption{Hybrid discrete adjoint pipeline. \emph{Top row:} the three outer steps (forward solve, iterative solve for $\vlam$, gradient extraction). The matrix-vector product $(\partial\R/\partial\vU)^\top\vv$ is computed by passing through the $M$ forward sub-steps in reverse and applying the three-step recipe (bottom box) at each sub-step, which reuses the efficient forward solver for the implicit half (1, 2) and one AD pass for the explicit half (3). Gradient extraction is a second pass through the chain with $\hat{\vU}_M = \vlam$. We use GMRES for the outer adjoint solve, but any matrix-free iterative method consuming this product works. The implementation differentiates the network once at the end of the gradient step rather than at every sub-step; the gradient is unchanged (\Cref{app:earsm_factoring}).}
\label{fig:adjoint_pipeline}
\end{figure}
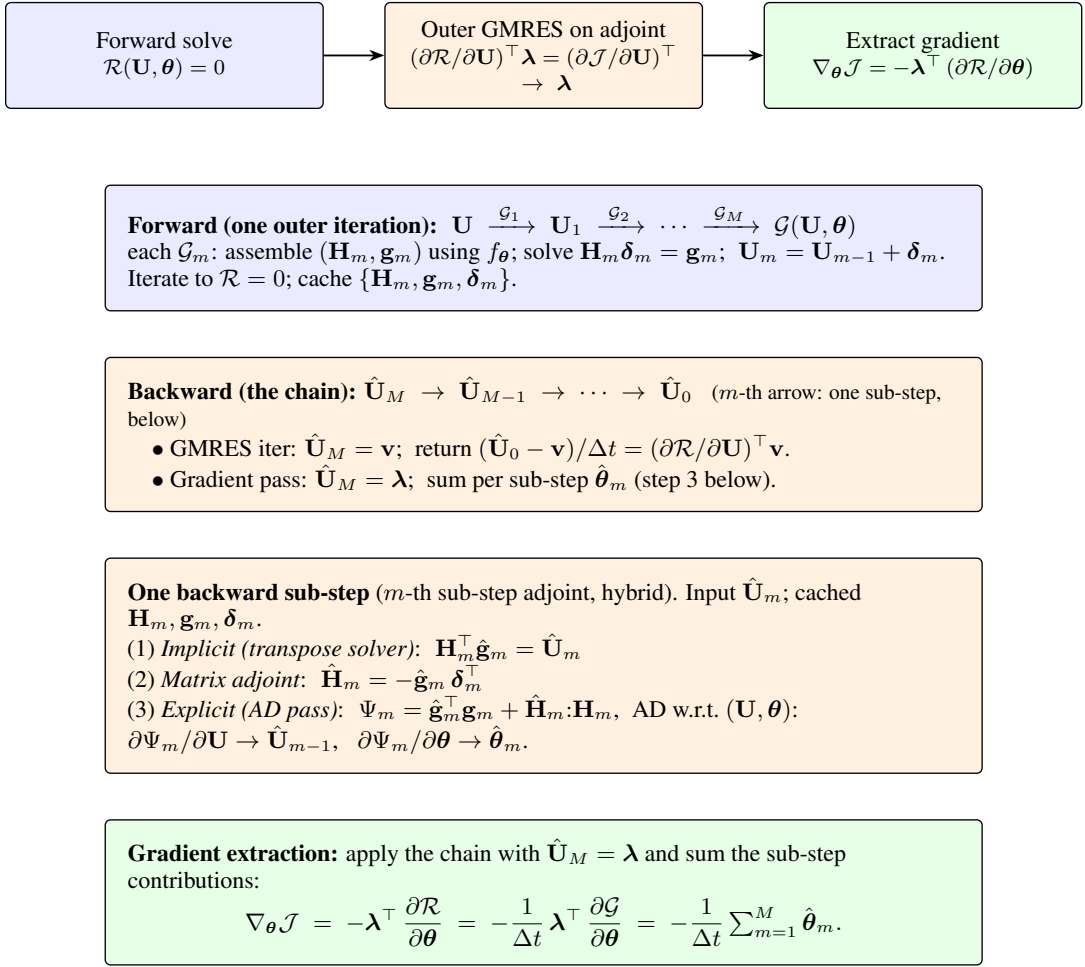

\Cref{app:adjoint} presents the full derivation and \Cref{fig:adjoint_pipeline} displays a schematic of the numerical method. Each forward sub-step solves an implicit linear equation
\begin{equation}
    \mathbf{H}_m(\vU_{m-1},\vtheta)\,\boldsymbol{\delta}_m = \mathbf{g}_m(\vU_{m-1},\vtheta),
    \qquad \vU_m = \vU_{m-1} + \boldsymbol{\delta}_m,
    \label{eq:HG}
\end{equation}
in which both the implicit operator $\mathbf{H}_m$ and the explicit right-hand side $\mathbf{g}_m$ depend on the state and, through the closure, on $\vtheta$. The structure of $\mathbf{H}_m$ is inherited from the PDE discretisation: tridiagonal for ADI sweeps, $2{\times}2$ block tridiagonal for the coupled $k$--$\omega$ block, symmetric Laplacian for the pressure projection. The forward step uses a fast solver for each (Thomas, block-Thomas, and BiCGSTAB respectively).

The adjoint of each sub-step has the same structure. Let $\hat{\vU}_m$ be the state adjoint coming from the next sub-step in the reverse chain of operations. The adjoint for each sub-step consists of (\Cref{fig:adjoint_pipeline}):
\begin{enumerate}[nosep]
    \item solve the adjoint of \eqref{eq:HG}, $\mathbf{H}_m^\top\hat{\mathbf{g}}_m = \hat{\vU}_m$, which is implicit with the same structure as the forward and uses the same numerical solver in transpose form;
    \item form the matrix adjoint $\hat{\mathbf{H}}_m = -\hat{\mathbf{g}}_m\,\boldsymbol{\delta}_m^\top$, which carries the dependence of the implicit operator on the state and the network;
    \item differentiate $\Psi_m = \hat{\mathbf{g}}_m^\top\mathbf{g}_m + \hat{\mathbf{H}}_m\!:\!\mathbf{H}_m$ with respect to $(\vU,\vtheta)$ by reverse-mode AD, returning $\hat{\vU}_{m-1}$ (the next sub-step's input) and $\hat{\vtheta}_m \coloneqq \partial\Psi_m/\partial\vtheta$ (the sub-step parameter contribution).
\end{enumerate}
We call this method a \emph{hybrid discrete adjoint} because step (1) reuses the forward numerical solver in transpose form, while step (3), which covers the assembly of $\mathbf{H}_m, \mathbf{g}_m$ from the state and parameters, including any embedded neural network, uses AD.

Applying steps (1)--(3) across all $M$ sub-steps in reverse and using the chain rule allows one to calculate the sequence of adjoints $\hat{\vU}_M \to \cdots \to \hat{\vU}_0$. The adjoint system \eqref{eq:adjoint_system} is then solved by any matrix-free iterative method that applies this chain. Convection makes the adjoint operator $(\partial\R/\partial\vU)^\top$ non-symmetric; stiff $k$--$\omega$ source terms make it ill-conditioned; and the pressure-Poisson sub-step couples it globally. These properties rule out conjugate gradients, and a direct solve is impractical at production grid sizes. We use GMRES, a matrix-free Krylov method suited to such operators; iterating the chain to a fixed point also converges, but it shares eigenvalues with the stiff forward solve and inherits its convergence rate. The product required by \eqref{eq:adjoint_system} is supplied by one pass of the chain: setting $\hat{\vU}_M = \vv$ returns $\hat{\vU}_0 = (\partial\mathcal{G}/\partial\vU)^\top\vv$, so $(\hat{\vU}_0 - \vv)/\Delta t = (\partial\R/\partial\vU)^\top\vv$ via $\R = (\mathcal{G}-\vU)/\Delta t$. Once $\vlam$ converges, we apply the chain once more with $\hat{\vU}_M = \vlam$, accumulating sub-step contributions:
\begin{equation}
    \nabla_{\vtheta}\J = -\frac{1}{\Delta t}\sum_{m=1}^{M}\hat{\vtheta}_m.
\end{equation}
In practice the chain rule is factored through the closure outputs so the network is differentiated only once at the end of the gradient pass (\Cref{app:earsm_factoring}); the gradient is unchanged. The method makes any semi-implicit PDE solver end-to-end differentiable in $\mathcal{O}(N)$ memory inside a standard autodiff framework, and the same code trains every network placement in this paper (closure coefficients, tensor-basis coefficients and a momentum source term). The full derivation and the validation of the adjoint against finite differences are in Appendix~\ref{app:adjoint}, and Appendix~\ref{app:cdr} applies the method to a second PDE family (convection--diffusion--reaction with a learned source term).

\paragraph{Empirical scaling of the hybrid adjoint.}\label{sec:scaling}

We argued above that unrolled backpropagation cannot scale on stiff systems; here we verify this empirically.
We compare three gradient calculation routes for grids $N\in\{16^2,32^2,64^2,128^2\}$:
(\textsc{Unroll}) reverse-mode AD through $K$ solver steps,
(\textsc{IFT-AD}) the implicit adjoint \eqref{eq:adjoint_system} with the linearised operator applied by autograd at every GMRES iteration, and (\textsc{IFT-Hybrid}) the same adjoint with the structured matrix-vector product of the hybrid adjoint (ours).
All three share a single converged steady state, computed once with the semi-implicit forward solver. The comparisons are run on CPU in double precision with a $128\,$GB memory ceiling (see Appendix~\ref{app:scaling} for a complete description).

\begin{table}[t]
\centering
\caption{Adjoint wall-time, peak memory, and iteration counts on grids of $N^2$ cells with a $128\,$GB memory budget. \textsc{Unroll} is run to depth $T$; on stiff grids its Neumann series diverges (\textbf{div}), on large grids it hits the memory ceiling (\textbf{mem}). \textsc{IFT-AD} uses GMRES on the full-AD Jacobian; \textbf{stall}: did not converge within $2000$ iterations, final relative residual in parentheses.}
\label{tab:scaling}
\small
\begin{tabular}{rlll}
\toprule
Grid    & \textsc{Unroll} ($T_{\max}$, time, mem) & \textsc{IFT-AD} (iters, time) & \textsc{IFT-Hybrid} (iters, time) \\
\midrule
$16^2$  & $2727$, $493$\,s, $34$\,GB             & $116$, $24$\,s                      & $106$, $\mathbf{11}$\,s  \\
$32^2$  & $3302$, $1073$\,s, $82$\,GB; \textbf{div}  & $552$, $226$\,s                   & $245$, $\mathbf{44}$\,s  \\
$64^2$  & -- , -- , $128$\,GB; \textbf{div}, \textbf{mem} & $2000$, $2022$\,s; \textbf{stall}\,($0.19$) & $365$, $\mathbf{143}$\,s \\
$128^2$ & -- , -- , $128$\,GB; \textbf{div}, \textbf{mem}    & $2000$, $3934$\,s; \textbf{stall}\,($0.98$) & $855$, $\mathbf{660}$\,s \\
\bottomrule
\end{tabular}
\end{table}

\begin{figure}[t]
    \centering
    \includegraphics[width=\textwidth]{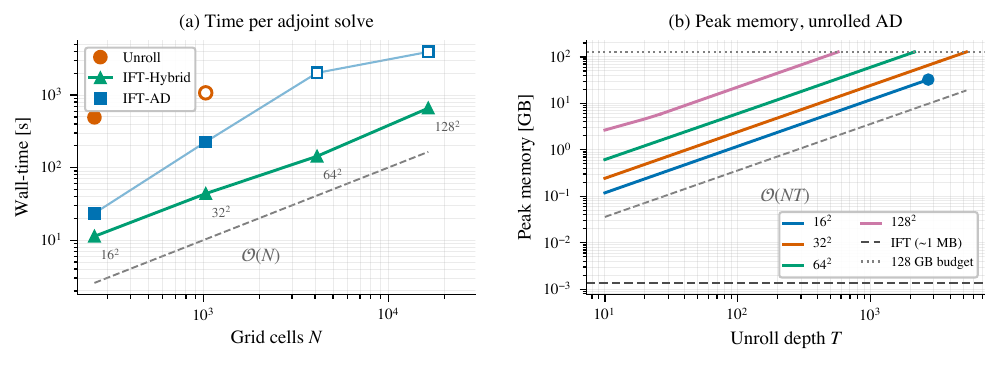}
    \caption{Gradient-computation scaling. \textbf{(a)}~Adjoint time vs.\ $N$. \textbf{(b)}~Peak memory: \textsc{Unroll} saturates the $128$\,GB budget. \textsc{IFT-Hybrid} is the only viable route to large-scale DARSM training.}
    \label{fig:scaling}
\end{figure}

\textsc{Unroll} is infeasible on two fronts (\Cref{tab:scaling}, \Cref{fig:scaling}).
On stiff grids its Neumann series diverges before the memory limit is reached, and memory grows as $\mathcal{O}(KN)$ and saturates the $128\,$GB budget at larger grids, limiting the reachable depth $T$.
The implicit adjoint avoids both failures by construction, solving the $T{\to}\infty$ limit in a single linear system with $\mathcal{O}(N)$ memory.
\textsc{IFT-AD} differentiates the inner iterative solves of a solver step as executed to finite tolerance, in the way a deep equilibrium model differentiates its layer~\citep{bai2019deq}, so the operator it returns is the transpose of an approximate solve rather than of the converged system. It is poorly conditioned, and GMRES stalls at $64^2$ and above.
\textsc{IFT-Hybrid} converges in under $900$ GMRES iterations at every grid tested and is the only method that produces a usable gradient at or above $64^2$, delivering a $2$--$5\times$ speed-up where \textsc{IFT-AD} succeeds.

Of the three routes, only the structured implicit adjoint scales, since unrolling is limited by memory and divergence and the full-autograd adjoint is poorly conditioned. It is the route by which DARSM-style closure learning can reach the problem sizes of engineering practice.

%% file: sections/experiments.tex
\section{Numerical Results for Canonical Turbulent Flows}
\label{sec:experiments}

We evaluate DARSM on two canonical benchmarks in turbulent fluid dynamics: fully-developed flow in a square duct and separated flow over periodic hills.
The closure, discretisation, and neural architecture are described in \Cref{sec:closure}; forward-solver correctness is verified against DNS and an independent OpenFOAM implementation in Appendix~\ref{app:forward_validation}.
The objective we minimise is the component-normalised velocity error
\begin{equation}
    \J(\vtheta) = \frac{1}{2}\sum_{i=1}^{3} \frac{w_i}{|\Omega|}
    \int_\Omega \bigl(U_i(\vx;\vtheta) - \overline{u_i^*}(\vx)\bigr)^{2}
    \, \mathrm{d}A,
    \qquad w_i = \|\overline{u_i^*}\|_\infty^{-2},
    \label{eq:Jvel}
\end{equation}

with $\overline{u_i^*}$ the time-averaged DNS target data and $w_i$ normalising the three velocity components to comparable scale.

\subsection{Comparison with alternative approaches}
\label{sec:comparison}

\begin{figure}[t]
    \centering
    \includegraphics[width=\textwidth]{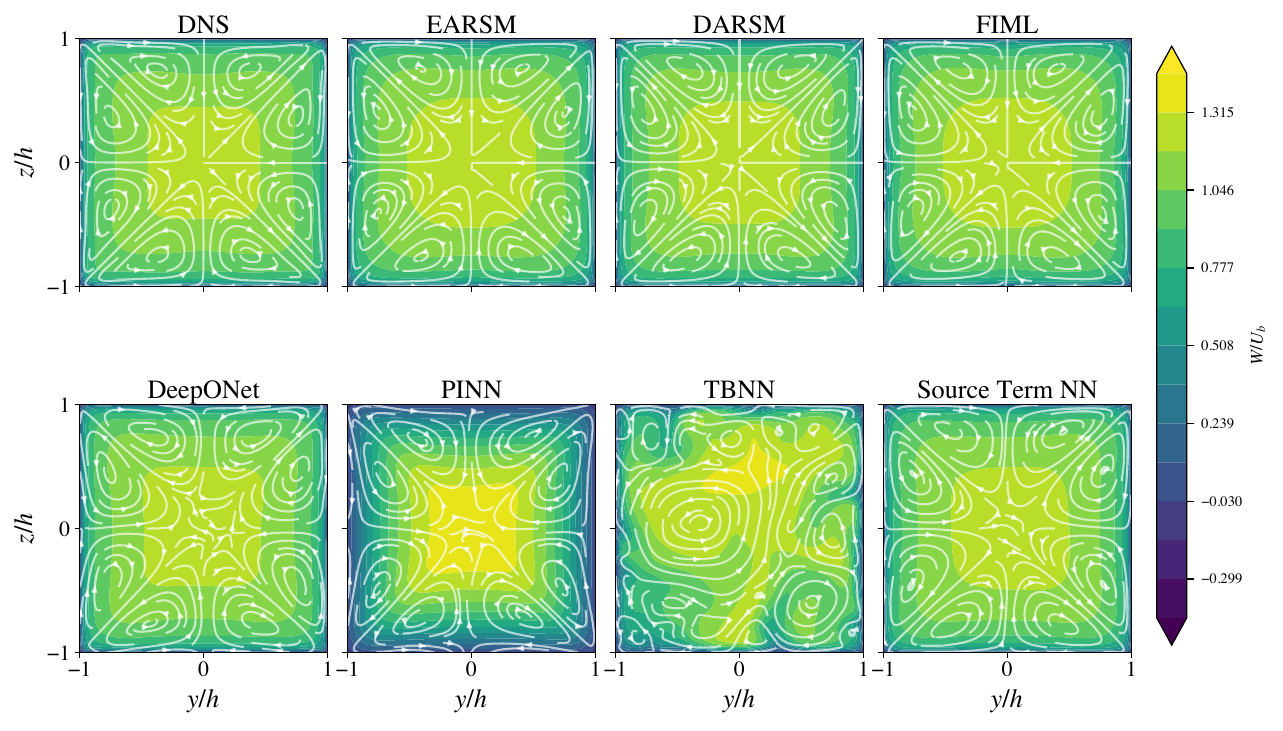}
    \caption{Streamwise velocity $W/U_b$ (filled contours) and in-plane secondary flow (white streamlines) at the test case $\Reb{=}11386$. DARSM recovers the symmetric eight-vortex secondary-flow topology of the DNS, whereas TBNN, PINN, DeepONet, FIML and the source-term closure produce distorted streamlines or weaker secondary motion.}
    \label{fig:duct_contours_main}
\end{figure}

The square-duct experiments use six DNS cases~\citep{pirozzoli2018turbulence,vinuesa2018turbulent} at $\Reb\in\{4410,5000,7000,11386,17800,40000\}$, symmetrised about the duct diagonals to remove finite-averaging noise.
We fix the training pair to $\{5000,17800\}$ and apply leave-one-out cross-validation over the four out-of-sample cases, each rotating as test while the remaining three drive early stopping and network capacity selection (full LOOCV protocol in Appendix~\ref{app:training}). We report the mean test error over folds, covering interpolation at $\Reb\in\{7000,11386\}$ and extrapolation both below and above the training range, with $\Reb=40000$ a factor of $2.3$ beyond the highest training case.

We compare against five approaches that do not train a closure end-to-end through the governing equations:
\emph{a-priori NN-EARSM} (the same network architecture as ours but trained offline on DNS data);
\emph{a-priori TBNN}~\citep{ling2016reynolds};
\emph{FIML}~\citep{parish2016paradigm} (per-case field inversion machine learning followed by supervised fit);
\emph{DeepONet}~\citep{lu2021learning} (a PDE-free neural operator surrogate); and a data-augmented physics-informed neural network \emph{PINN}~\citep{raissi2019physics}. We also compare against one approach that does. This \emph{source-term} closure replaces the physics-derived structure by a learned momentum body force in \Cref{eq:rans_main}, the form used by neural-correction methods~\citep{kochkov2021machine,sirignano2020dpm}, supplied by a network with DARSM's architecture and trained with the identical objective, optimiser, and adjoint.
Architectures, hyperparameters, and training protocols for all baselines are described in Appendix~\ref{app:baselines}.

\begin{table}[t]
\centering
\caption{Square-duct velocity error $\J \times 10^{3}$ on the Reynolds-number split. ``Div.''~= forward solver diverged when the trained closure was inserted. A-priori models are trained offline on DNS data; the TBNN cannot be trained end-to-end through the solver (see text and Appendix~\ref{app:tbnn_e2e}). Lower error is better. DARSM values are mean$\pm$std over seeds.}
\label{tab:duct}
\small
\begin{tabular}{lccc}
\toprule
 & $\J^\text{train}$ & $\J^\text{test}$ & $\Reb{=}40000$ \\
\midrule
Default EARSM (baseline)~\citep{wallin2000explicit}  & $7.53$ & $7.38$ & $8.97$ \\
A-priori NN-EARSM & Div. & Div. & Div. \\
A-priori TBNN (frozen)~\citep{ling2016reynolds}    & $584$ & $609$ & $944$ \\
FIML~\citep{parish2016paradigm}           & $28.1$ & $17.4$ & $8.56$ \\
DeepONet~\citep{lu2021learning}           & $7.06$ & $7.92$ & $7.55$ \\
PINN (data augmented)~\citep{raissi2019physics}            & $40.1$ & $57.1$ & $109.0$ \\
Source-term NN (adjoint-trained)            & $4.14$ & $5.17$ & $6.83$ \\
\textbf{DARSM}                    & $\mathbf{2.45\pm0.10}$ & $\mathbf{3.56\pm0.27}$ & $\mathbf{3.75\pm0.64}$ \\
\bottomrule
\end{tabular}
\end{table}

In mean out-of-sample error, the two adjoint-trained closures (DARSM and the source-term closure) are the only methods that improve on the baseline on the duct (\Cref{tab:duct}, \Cref{fig:duct_contours_main}). DARSM outperforms the source-term closure by $\sim30\%$ on the test split although both share the same end-to-end training, and it is the only method that also improves on the out-of-distribution periodic hills (\Cref{sec:hills}). Each offline or solver-free alternative underperforms through a distinct structural cause.

FIML isolates the structural mechanism. Its field-inversion step reaches $\J^\text{train}{=}3.96\times 10^{-3}$ with solver-in-the-loop optimisation, showing that a good closure exists and is discoverable, but once that coefficient field is refit by supervised regression from local invariants (Step~2), accuracy degrades by an order of magnitude.
The cause is distribution shift (\Cref{sec:related}). The network is trained on invariants at the field-inversion-converged RANS state but is queried at deployment on the state the solver produces with the imperfect network, which sits in a different region of feature space. The same effect has been reported in other settings: \citet{um2020solver} showed that identical CNN correction architectures reduce coarse-grid errors by $73\%$ when trained through the solver versus the offline-trained baseline in unsteady wake flow, a gap attributable entirely to training-time distribution shift.

\begin{figure}[t]
    \centering
    \includegraphics[width=\textwidth]{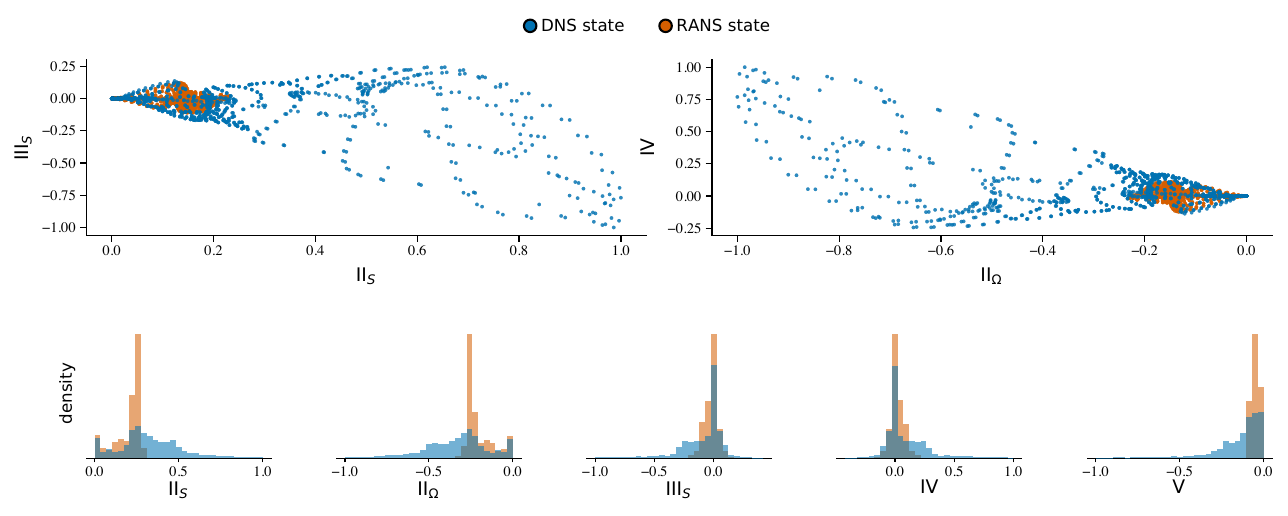}
\caption{Distribution shift in the $(\mathbf{S},\boldsymbol\Omega)$-invariant
feature space the a-priori network consumes. All invariants normalised
per case ($\tau{=}1/\omega_\text{RANS}$). \textbf{Top}: joint distributions
at $\Reb{=}5000$ of the strain pair $(\mathrm{II}_S,\mathrm{III}_S)$ and the
rotation/coupling pair $(\mathrm{II}_\Omega,\mathrm{IV})$, at the DNS state
(training) and the converged default RANS state (deployment). \textbf{Bottom}:
1-D marginals of all five invariants pooled across
$\Reb\in\{5000, 7000, 11386, 40000\}$. Every feature is displaced between
training and deployment, the covariate shift whose consequence is
quantified in \Cref{fig:shift}.}
\label{fig:shift_features}
\end{figure}

\begin{figure}[t]
    \centering
    \includegraphics[width=\textwidth]{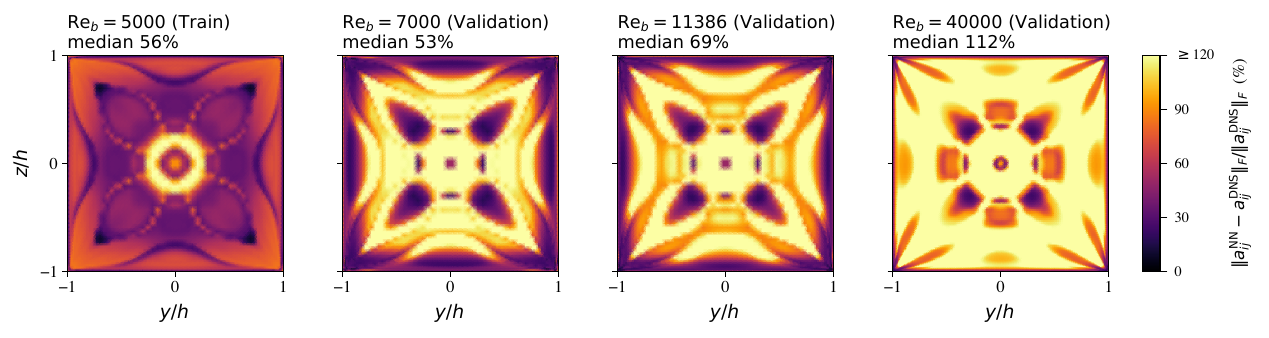}
    \caption{Pointwise relative Frobenius error of the a-priori TBNN anisotropy~\citep{ling2016reynolds} versus DNS, evaluated on RANS-state invariants at four Reynolds numbers (one in training, three held out). The closure was trained offline on DNS-state invariants but queried at deployment on RANS-state invariants (\Cref{fig:shift_features}); $\sim50\%$ median error appears even at the training $\Reb=5000$ and persists across all $\Reb$, concentrating at corners and walls.}

    \label{fig:shift}
\end{figure}

For the a-priori TBNN the shift is visible directly in feature space, where every invariant the network consumes is displaced between the DNS states it was trained on and the RANS states it is queried on (\Cref{fig:shift_features}). \Cref{fig:shift} shows the consequence at deployment, where the closure is $50$--$120\%$ wrong in Frobenius norm at every case in the sweep. These mispredictions compound with the intrinsic ill-conditioning of the RANS equations when Reynolds stresses are specified a priori. On the same square-duct and periodic-hill geometries, \citet{wu2019rans} show that even $0.5\%$ perturbations of the Reynolds stresses can produce $\mathcal{O}(10\%)$ velocity errors at the Reynolds numbers we consider.
Training through the solver would remove the shift, since the network is then queried on the solver's own states, but the TBNN cannot be trained this way. We trained it end-to-end with the same objective, optimiser, and adjoint as DARSM. From a small-output initialisation the network gives $C_\mu\approx0$, which collapses the effective eddy viscosity, and the first forward solve diverges at every network width. Initialised at the Boussinesq limit, training starts, but the error stays well above DARSM's and the optimiser soon leaves the region in which the forward solver is stable. The tensor basis is a completeness result that contains no physics, so the network can represent stresses that act as a negative viscosity as easily as physical ones, and nothing in the loss penalises solver instability (Appendix~\ref{app:tbnn_e2e}). The TBNN in \Cref{tab:duct} is therefore the offline network, with its anisotropy frozen at the converged default state (Appendix~\ref{app:baseline_apriori}).

DeepONet sidesteps the closure problem altogether, mapping inputs directly to flow fields, but with only two Reynolds numbers it has insufficient data for extrapolation. PINNs include the PDE as a penalty in the loss, which is a natural fit for inverse problems and data assimilation; however, on this problem PINN training does not converge, even with multi-case sharing, because spectral bias suppresses the multi-scale spatial structure (sharp near-wall gradients and the $\mathcal{O}(1\%)$ secondary motion that must be resolved against a much larger streamwise flow), and the multi-term loss exhibits gradient pathologies from the nonlinear $k$--$\omega$ source coupling~\citep{krishnapriyan2021, wang2022,ji2021}.

\subsection{Square duct}
\label{sec:duct}

Trained on only two Reynolds numbers, the adjoint-optimized DARSM closure generalises across an order of magnitude of $\Reb$, reducing the velocity error by a factor of two at the extrapolation endpoint $\Reb{=}40000$ ($2.3\times$ the highest Reynolds number training case) as compared to the default RANS-EARSM model. DARSM, when trained on only two aspect ratios, also accurately extrapolates to wider ducts with a threefold reduction in error. \Cref{tab:duct_main} reports per-case results.

\begin{table}[t]
\centering
\caption{Square-duct velocity error $\J\times 10^{3}$. \emph{Top}: Reynolds-number generalisation, training on $\Reb{\in}\{5000,17800\}$ with the remaining four cases as test. \emph{Bottom}: aspect-ratio generalisation, training on $\{\mathrm{AR}{=}1,\mathrm{AR}{=}3\}$ with $\{\mathrm{AR}{=}5,\mathrm{AR}{=}7\}$ out-of-sample. DARSM values are mean$\pm$std over seeds.}
\label{tab:duct_main}
\setlength{\tabcolsep}{6pt}
\renewcommand{\cellalign}{cc}
\begin{tabular}{p{3.3cm}cccccc|c}
\toprule
 & \multicolumn{2}{c}{\textsc{train}} & \multicolumn{4}{c|}{\textsc{test}} & \textsc{test mean} \\
\cmidrule(lr){2-3}\cmidrule(lr){4-7}
\multicolumn{8}{l}{\emph{Reynolds number (AR=1)}} \\
 & $\Reb{=}5000$ & $17800$ & $4410$ & $7000$ & $11386$ & $40000$ &  \\
Default EARSM & $8.80$ & $6.26$ & $7.95$ & $6.59$ & $6.01$ & $8.97$ & $7.38$ \\
\textbf{DARSM}
 & \makecell{$\mathbf{2.70}$\\\scriptsize$\pm0.28$}
 & \makecell{$\mathbf{2.20}$\\\scriptsize$\pm0.09$}
 & \makecell{$\mathbf{2.71}$\\\scriptsize$\pm0.04$}
 & \makecell{$\mathbf{4.30}$\\\scriptsize$\pm0.33$}
 & \makecell{$\mathbf{3.49}$\\\scriptsize$\pm0.14$}
 & \makecell{$\mathbf{3.75}$\\\scriptsize$\pm0.64$}
 & \makecell{$\mathbf{3.56}$\\\scriptsize$\pm0.27$} \\
\midrule
\multicolumn{8}{l}{\emph{Aspect ratio (Re$_\tau{=}180$)}} \\
 & $\mathrm{AR}{=}1$ & $3$ & $5$ & $7$ & & & \\
Default EARSM & $8.86$ & $7.16$ & $6.57$ & $5.22$ & & & $5.90$ \\
\textbf{DARSM}
 & \makecell{$\mathbf{3.98}$\\\scriptsize$\pm1.56$}
 & \makecell{$\mathbf{2.63}$\\\scriptsize$\pm0.77$}
 & \makecell{$\mathbf{1.94}$\\\scriptsize$\pm0.61$}
 & \makecell{$\mathbf{1.46}$\\\scriptsize$\pm0.31$}
 & &
 & \makecell{$\mathbf{1.70}$\\\scriptsize$\pm0.46$} \\
\bottomrule
\end{tabular}
\end{table}

\paragraph{Reynolds-number generalisation.}
The closure achieves $\J^\text{test}{=}3.56\times 10^{-3}$, a $52\%$ reduction in squared velocity error over the baseline, with consistent gains across interpolation and extrapolation. The improvement extends to the Reynolds stresses, with a lower error in the shear-stress components of the anisotropy at every Reynolds number (Appendix~\ref{app:app_metrics}).

\begin{figure}[t]
    \centering
    \includegraphics[width=\textwidth]{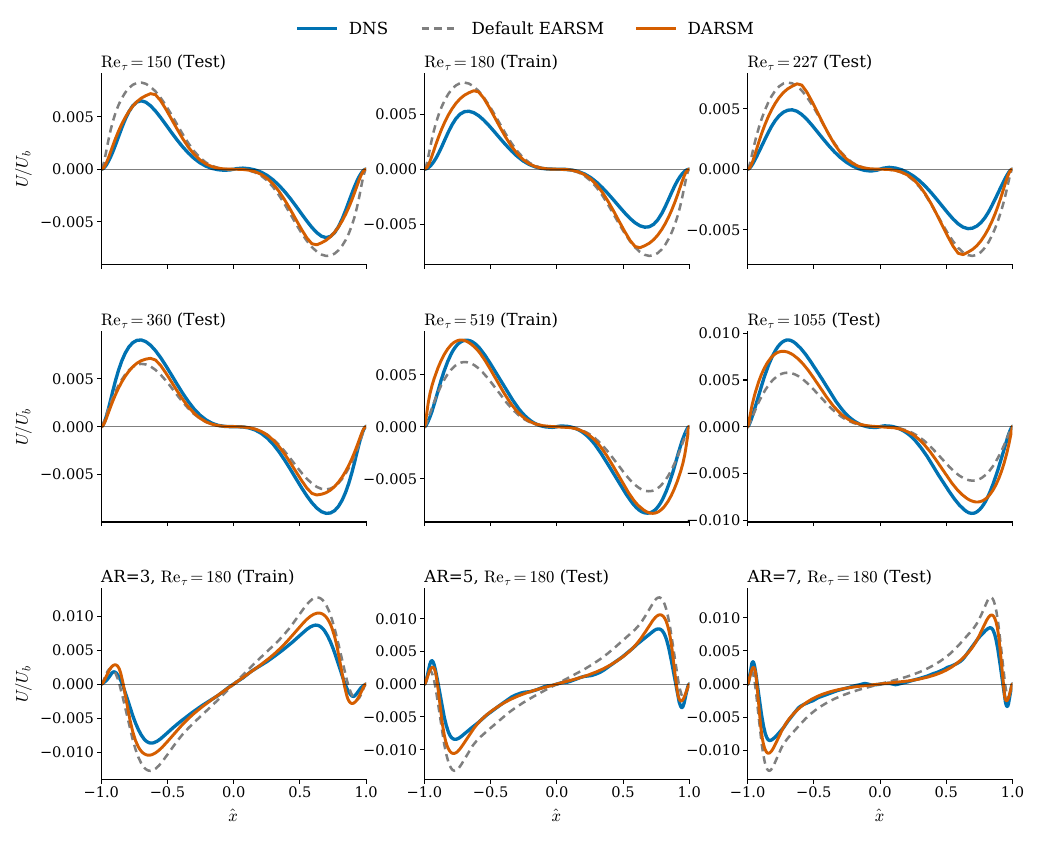}
    \caption{Cross-stream velocity $U/U_b$ along the horizontal midplane $y{=}L_y/2$ versus $\hat{x}{=}2(x-L_x/2)/L_x{\in}[{-}1,1]$. Top two rows: six $\Reb$ cases at $\mathrm{AR}{=}1$ (train: $\Reb{\in}\{5000,17800\}$); panels are labelled by friction Reynolds number, with $\mathrm{Re}_\tau{=}150, 180, 227, 360, 519, 1055$ corresponding to $\Reb{=}4410, 5000, 7000, 11386, 17800, 40000$. Bottom row: three AR
cases at $\mathrm{Re}_\tau{=}180$ (train: $\mathrm{AR}{\in}\{1,3\}$). The default EARSM systematically mispredicts secondary-flow amplitude; the
adjoint-trained closure recovers amplitude and sign across all cases, including extrapolation to $\Reb{=}40000$ (top right) and $\mathrm{AR}{=}7$ (bottom right).}
    \label{fig:profiles}
\end{figure}

\Cref{fig:profiles} compares the cross-stream velocity $U$ along the horizontal midplane ($y{=}L_y/2$) against DNS at every Reynolds-number and aspect-ratio case. Both the default closure and the adjoint-trained closure reproduce the qualitative eight-vortex secondary-flow structure, but the default closure systematically mispredicts its amplitude; the adjoint-trained closure recovers both amplitude and sign reversal at all training and test cases, including extrapolation to $\Reb{=}40000$ and to the wider out-of-sample ducts ($\mathrm{AR}{\in}\{5,7\}$).

\paragraph{Aspect-ratio generalisation.}
Training on $\{\mathrm{AR}{=}1,\mathrm{AR}{=}3\}$ with $\{\mathrm{AR}{=}5,\mathrm{AR}{=}7\}$ as test, DARSM reduces the test error from $5.90\times 10^{-3}$ to $1.70\times 10^{-3}$ as compared to the default RANS model, which is a $3\times$ improvement for pure extrapolation (both test aspect ratios are wider than any training case).

\subsection{Periodic hills}
\label{sec:hills}

\paragraph{Shape generalisation.}
Trained on a single hill geometry, the closure transfers to hill shapes it has never seen and reduces the velocity error by a factor of $4$ in the mean.
Periodic-hills flow at $\Reb{=}5600$ has a hill-length parameter $\alpha$ that strongly influences the separation-bubble length and reattachment point; we train on $\alpha{=}1.0$ and evaluate on $\alpha\in\{0.5, 0.75, 1.25, 1.5\}$, spanning half to one-and-a-half times the training hill. The DNS target at $\alpha{=}1.0$ is from~\citet{breuer2009flow} and the DNS at the out-of-sample $\alpha$ values from the parametric-hills database of~\citet{xiao2020flows}.

\begin{table}[t]
\centering
\caption{Periodic hills at $\Reb{=}5600$: velocity error $\J \times 10^{3}$ on five hill shapes; lower is better. Closures trained on the duct are deployed without retraining, so every shape is out of sample and the mean is over all five; $^{\dagger}$mean over the four non-training shapes. FIML (inversion) is the per-geometry field-inversion step against each geometry's own DNS and is in sample. Protocols in Appendix~\ref{app:baselines_hills}; Div.~= solver diverged; DARSM values are mean$\pm$std over seeds.}
\label{tab:hills}
\footnotesize
\setlength{\tabcolsep}{3pt}
\begin{tabular}{lccccc|c}
\toprule
 & $\alpha{=}1.0$ & $\alpha{=}0.5$ & $\alpha{=}0.75$ & $\alpha{=}1.25$ & $\alpha{=}1.5$ & Mean \\
\midrule
Default EARSM & $5.65$ & $1.73$ & $2.62$ & $8.18$ & $10.49$ & $5.73$ \\
\midrule
\multicolumn{7}{l}{\emph{Trained on hills ($\alpha{=}1.0$)}} \\
\textbf{DARSM}  & $\mathbf{0.29}{\scriptstyle\pm0.01}$ & $\mathbf{3.06}{\scriptstyle\pm0.20}$ & $\mathbf{1.29}{\scriptstyle\pm0.08}$ & $\mathbf{0.65}{\scriptstyle\pm0.09}$ & $\mathbf{1.02}{\scriptstyle\pm0.16}$ & $\mathbf{1.50}{\scriptstyle\pm0.13}^{\dagger}$ \\
\midrule
\multicolumn{7}{l}{\emph{Trained on duct, deployed without retraining}} \\
\textbf{DARSM}  & $\mathbf{0.75}$ & $\mathbf{3.01}$ & $\mathbf{1.32}$ & $\mathbf{0.53}$ & $\mathbf{0.58}$ & $\mathbf{1.24}$ \\
DeepONet        & $112.1$ & $128.5$ & $123.3$ & $106.1$ & $103.4$ & $114.7$ \\
Source-term NN  & $46.5$ & $25.6$ & $46.3$ & $56.6$ & $68.4$ & $48.7$ \\
FIML            & $73.6$ & $22.8$ & $15.2$ & $153$ & $108$ & $74.5$ \\
\midrule
\multicolumn{7}{l}{\emph{Offline-trained and solver-free}} \\
A-priori NN-EARSM & $12.3$ & $10.4$ & $9.1$ & $22.7$ & $24.7$ & $15.8$ \\
A-priori TBNN (frozen) & $105$ & $83$ & $199$ & $262$ & $459$ & $222$ \\
FIML (inversion, in sample) & $3.4$ & $1.1$ & $1.5$ & $4.9$ & $6.8$ & $3.5$ \\
PINN            & Div. & Div. & Div. & Div. & Div. & -- \\
\bottomrule
\end{tabular}
\end{table}

\begin{figure}[t]
    \centering
    \includegraphics[width=\textwidth]{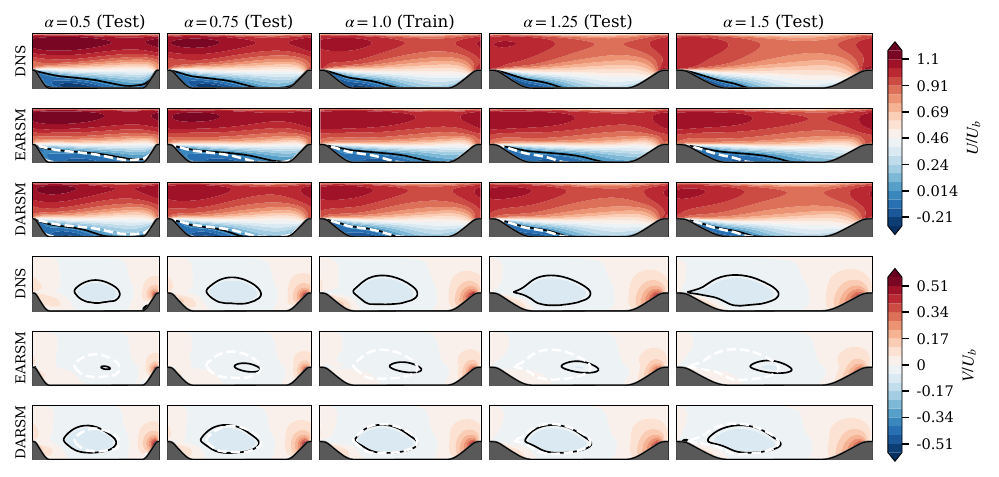}
\caption{Periodic-hills cross-sections at $\Reb{=}5600$ for five hill shapes
$\alpha\in\{0.5, 0.75, 1.0, 1.25, 1.5\}$ (columns); the training case is
$\alpha{=}1.0$, the rest are test. Rows 1--3: streamwise velocity $U/U_b$ (DNS, default EARSM, DARSM). Rows 4--6: wall-normal velocity $V/U_b$ (same sources). Solid black contours mark each model's $U,V{=}0$ lines; dashed white re-draws the DNS reference on the model rows.}\label{fig:hills}
\end{figure}

\Cref{tab:hills} (\emph{trained on hills}) gives per-$\alpha$ results. The out-of-sample mean $\J$ falls well below the baseline, with the largest gain at $\alpha{=}1.25$ ($12\times$ reduction). Performance degrades at $\alpha=0.5$ ($1.8\times$ the baseline error), the geometry furthest from training in both hill length and bubble extent.
\Cref{fig:hills} shows the streamwise and wall-normal velocity fields across all five hill shapes. The default EARSM closure systematically overpredicts the recirculation bubble, whose $U{=}0$ contour (solid black) never matches the DNS target (dashed white), and the wall-normal $V$ field loses the upwash over the hill crest. DARSM recovers the bubble shape and the upwash structure of $V$ across most out-of-sample geometries, including the long $\alpha=1.5$ hill; agreement with the separation bubble degrades at $\alpha=0.5$, though the closure still improves on $V$. Separation, reattachment, and recirculation are governed by the anisotropic stress that the linear eddy-viscosity hypothesis cannot represent; the EARSM's tensor structure provides the right functional form, and the adjoint provides the coefficients that make it accurate across out-of-sample hill geometries.

\paragraph{Separation, reattachment, and turbulent kinetic energy.}
The reduction in velocity error carries over to the quantities of engineering interest in a separated flow (\Cref{tab:hills_metrics} in Appendix~\ref{app:app_metrics}). The separation onset matches the DNS at every geometry. The default closure places the reattachment point several hill heights too far downstream, which DARSM corrects on all geometries except $\alpha{=}0.5$, where it places reattachment slightly upstream of the DNS, the same geometry on which the velocity error degrades. The turbulent-kinetic-energy error, which the objective does not see, falls at every geometry.

\paragraph{Cross-regime generalisation.}
We deploy the duct-trained closure from \Cref{sec:duct} directly on the periodic-hills solver, without retraining (\Cref{tab:hills}, \emph{trained on duct}). It reduces the error over the default EARSM by $4.6\times$ in the mean over all five shapes, and on the four shapes shared with the hills-trained model it is on par with that model, despite the training data containing no separation or separated shear-layer physics. None of the other duct-trained baselines transfers. DeepONet, FIML and the source-term closure are one to two orders of magnitude worse than the default EARSM, and since the source-term closure was trained with the same adjoint as DARSM, end-to-end training alone does not make a learned correction generalise. The offline-trained closures fail on the hills as on the duct, so only DARSM improves on the default closure out of sample. The learned parameters stay meaningful on a flow whose physics, dominated by separation and reattachment, is fundamentally different from the corner-driven secondary flow of the training geometry, because they enter through DARSM's physics-derived algebraic structure. Appendix~\ref{app:coeff_corrections} shows which coefficient corrections the duct- and hills-trained networks share, and that fixing the constants at those mean corrections recovers only part of the improvement.

%% file: sections/discussion.tex
\section{Discussion}
\label{sec:discussion}

\paragraph{Data-efficient generalisation.}
Adjoint-optimized DARSM accurately generalises from small datasets, with models trained on only one or two flow cases reducing the mean out-of-sample velocity error against the baseline across all three generalisation regimes (Reynolds number, duct aspect ratio, periodic-hill geometry). DARSM learns from so few flows because most of the closure is fixed by physics. The EARSM's tensor structure comes from the Reynolds stress transport equations under weak equilibrium, and the network only adjusts its empirical constants as functions of the local invariants, the same quantities on which the EARSM coefficients depend. Every correction therefore reaches the anisotropy through the EARSM's algebra, and the network never has to learn the input-to-anisotropy map from scratch. What it learns is a set of interpretable model parameters rather than the stress itself, whose form changes from one flow to the next. Training through the converged forward solver, which encodes conservation, boundary conditions, and field coupling, keeps those parameters consistent with the governing equations. Offline training of the same network on the same data crashes the solver on the duct and degrades accuracy on the hills. FIML's two-step design shows the same contrast within a single method. Its solver-in-the-loop field inversion (step~1) finds a coefficient field that brings the RANS solution close to the DNS target, but the offline supervised regression that refits this field from local invariants (step~2) degrades accuracy by an order of magnitude.

\paragraph{Transfer across flow regimes.}
The duct-to-hills transfer tests this design most directly. A closure trained only on square-duct data, where the dominant anisotropy mechanism is corner-driven secondary motion, transfers directly to periodic hills, where the dominant mechanism is separated shear-layer anisotropy, reducing velocity error by $4.6\times$ over the classical baseline and matching the accuracy of a closure trained on hills data. The corrections transfer because they act on the ARSM's coefficients through invariants that both flows share. A closure that replaced the EARSM form with a learned black-box regression, or that learned corrections directly in velocity space, would be unlikely to generalise in this way; DeepONet, FIML and the source-term closure trained on the same duct data are one to two orders of magnitude worse on the periodic hills.

\paragraph{The governing equations as physics-based structure for learning.}
The alternative methods we benchmark each omit one of the structural ingredients of DARSM, so together they test its two parts separately. Offline-trained closures (TBNN, a-priori NN-EARSM, and FIML's supervised step) keep their parameterisation but train the network on states the solver never sees, and incur distribution shift at deployment. Operator surrogates (DeepONet) and PDE-residual learners (PINNs) avoid the PDE solver entirely, forgoing the conservation, boundary condition, and coupling structure it enforces. The neural source-term baseline trains through the solver but leaves the closure form unconstrained, so the learned body force has no physical structure to remain valid outside the training distribution, and on the hills its error is an order of magnitude above the default closure. The tensor-basis network cannot be trained through the solver at all, because without a physics-derived structure its stress can act as a negative viscosity (\Cref{sec:comparison}). DARSM is the only method in our benchmark that combines physics-derived closure structure, invariant-aligned coordinates, and adjoint training through the solver.

\paragraph{Limitations.}

The constant-memory steady-state adjoint formulation requires the forward solver to converge to a steady state, consistent with RANS's premise that the time-averaged mean of the (unsteady) flow is stationary. Genuinely unsteady flows where no time-stationary mean exists are better treated by scale-resolving methods such as LES, for which unsteady adjoint formulations have been developed \citep{sirignano2020dpm}. All results are on 2D grids. The closure and adjoint formulations are dimension-agnostic and the $\mathcal{O}(N)$ scaling of \Cref{tab:scaling} carries to 3D, where the implicit sub-steps remain tridiagonal and block-tridiagonal solves plus one Poisson solve, but generalisation on genuinely three-dimensional flows, with three-dimensional separation, swirl, and wall-curvature interactions, requires empirical evaluation beyond this work. We benchmark on two canonical flows widely used in turbulence modelling, the square duct and periodic hills. Other configurations such as free shear layers (jets, mixing layers), external aerodynamics (aerofoils), and compressible flows were not tested and may behave differently. The weak-equilibrium assumption underlying the EARSM has known limits (strong streamline curvature, rapid acceleration, substantial non-equilibrium dynamics), which any closure built on it, including DARSM, inherits. Like any data-driven closure, DARSM also requires a source of high-fidelity reference data for training.

\section{Conclusion}
\label{sec:conclusion}

DARSM is a deep learning closure for RANS in which neural networks model the empirical constants of an algebraic Reynolds stress equation derived under weak equilibrium. The model is trained by optimizing over the entire PDE system using discrete semi-implicit adjoint equations. The construction is solver-agnostic and extends to any semi-implicit PDE solver, suggesting a general path for embedding neural-network components in physics-derived models trained end-to-end through the governing equations. Using only limited training data, DARSM improves upon the classical EARSM on two widely-used benchmarks for turbulence models, the square duct and periodic hills. Trained on one or two cases, it generalises across Reynolds number, duct aspect ratios, periodic-hill geometry and flow regime, including a transfer from attached duct flow to separated hill flow without retraining; it reduces out-of-sample velocity error by $2$--$4\times$ in the mean and up to $12\times$ per case, and outperforms the six ML methods benchmarked. These results rest on combining a closure whose tensor structure and invariant basis are derived from first principles with a steady-state adjoint that provides solver-consistent gradients at $\mathcal{O}(N)$ memory, and the benchmarks show that neither suffices alone. Once trained, DARSM runs at the cost of a standard RANS calculation, and three-dimensional engineering flows are the next step.

\paragraph{Reproducibility and broader impact.}
The DNS reference datasets are publicly available from their original sources. The trained DARSM parameters, the forward solver, the adjoint, and the baseline training pipelines will be released at \url{https://github.com/danieldeh/DARSM}. Training protocols, hyperparameters, and the verification of the forward solver and adjoint are given in Appendices~\ref{app:forward_validation}, \ref{app:vjp}, \ref{app:training_procedure}, \ref{app:ablations} and~\ref{app:baselines}. We foresee no negative societal impacts specific to this contribution.

%% file: sections/appendix_arxiv.tex
\section{Physics-derived closure}
\label{app:earsm}

By the Cayley--Hamilton theorem, any symmetric deviatoric tensor can be
written as a linear combination of ten independent tensor groups formed
from the normalised strain $\mathbf{S}^{*}$ and rotation
$\boldsymbol{\Omega}^{*}$~\citep{pope1975more}:
\begin{equation}\label{eq:popebasis}
    a_{ij} = \sum_{n=1}^{10} \beta^{(n)}\,\hat{T}^{(n)}_{ij},
\end{equation}
where the Pope basis tensors are~\citep{wallin2000explicit}
\begin{equation}\label{eq:pope_tensors}
\begin{aligned}
    \hat{T}^{(1)} &= \mathbf{S}^{*}, &
    \hat{T}^{(2)} &= \mathbf{S}^{*2} - \tfrac{1}{3}\mathrm{II}_S\,\mathbf{I}, \\
    \hat{T}^{(3)} &= \boldsymbol{\Omega}^{*2} - \tfrac{1}{3}\mathrm{II}_\Omega\,\mathbf{I}, &
    \hat{T}^{(4)} &= \mathbf{S}^{*}\boldsymbol{\Omega}^{*} - \boldsymbol{\Omega}^{*}\mathbf{S}^{*}, \\
    \hat{T}^{(5)} &= \mathbf{S}^{*2}\boldsymbol{\Omega}^{*} - \boldsymbol{\Omega}^{*}\mathbf{S}^{*2}, &
    \hat{T}^{(6)} &= \mathbf{S}^{*}\boldsymbol{\Omega}^{*2} + \boldsymbol{\Omega}^{*2}\mathbf{S}^{*} - \tfrac{2}{3}\mathrm{IV}\,\mathbf{I}, \\
    \hat{T}^{(7)} &= \mathbf{S}^{*2}\boldsymbol{\Omega}^{*2} + \boldsymbol{\Omega}^{*2}\mathbf{S}^{*2} - \tfrac{2}{3}\mathrm{V}\,\mathbf{I}, &
    \hat{T}^{(8)} &= \mathbf{S}^{*}\boldsymbol{\Omega}^{*}\mathbf{S}^{*2} - \mathbf{S}^{*2}\boldsymbol{\Omega}^{*}\mathbf{S}^{*}, \\
    \hat{T}^{(9)} &= \boldsymbol{\Omega}^{*}\mathbf{S}^{*}\boldsymbol{\Omega}^{*2} - \boldsymbol{\Omega}^{*2}\mathbf{S}^{*}\boldsymbol{\Omega}^{*}, &
    \hat{T}^{(10)} &= \boldsymbol{\Omega}^{*}\mathbf{S}^{*2}\boldsymbol{\Omega}^{*2} - \boldsymbol{\Omega}^{*2}\mathbf{S}^{*2}\boldsymbol{\Omega}^{*},
\end{aligned}
\end{equation}
and the scalar coefficients $\beta^{(n)}$ depend on at most the five invariants
\begin{equation}\label{eq:pope-inv}
    \mathrm{II}_S = \tr\mathbf{S}^{*2}, \quad
    \mathrm{II}_\Omega = \tr\boldsymbol{\Omega}^{*2}, \quad
    \mathrm{III}_S = \tr\mathbf{S}^{*3}, \quad
    \mathrm{IV} = \tr(\mathbf{S}^{*}\boldsymbol{\Omega}^{*2}), \quad
    \mathrm{V} = \tr(\mathbf{S}^{*2}\boldsymbol{\Omega}^{*2}).
\end{equation}
This decomposition is purely mathematical: any frame-invariant closure can be written in this form. The design rationale for DARSM's choice of embedding, which lets the network perturb the empirical constants of an explicit ARSM closure rather than learn the $\beta^{(n)}$ directly, is given in \Cref{sec:rationale}.

Variants of the linear eddy-viscosity hypothesis~\citep{craft1996development} have introduced invariant dependence by prescribing $C_\mu(\mathrm{II}_S, \mathrm{II}_\Omega)$ through a realizability-motivated damping function. This acknowledges that invariant-dependent sensitivities are needed, but the functional form is a modelling choice added to repair specific failure modes rather than a derivation from the transport physics underlying the ARSM.

Below we derive the implicit ARSM equation that determines the $\beta^{(n)}$, give the explicit Wallin--Johansson solution, and present the stress splitting used by the forward solver.

\subsection{Explicit algebraic Reynolds stress model (EARSM)}
\label{app:earsm_derivation}
The transport equation for the Reynolds stress is~\citep{pope2000turbulent}
\begin{equation}\label{eq:rst}
    \frac{\mathrm{D}\langle u_i' u_j'\rangle}{\mathrm{D}t}
    - \frac{\partial T_{ijl}}{\partial x_l}
    = \mathcal{P}_{ij} - \varepsilon_{ij} + \Pi_{ij},
\end{equation}
where $T_{ijl}$ is the transport flux of the Reynolds stress, $\mathcal{P}_{ij} = -\langle u_i' u_k'\rangle U_{j,k}
- \langle u_j' u_k'\rangle U_{i,k}$ is turbulent production, $\varepsilon_{ij}$ is the
dissipation rate tensor, and $\Pi_{ij}$ is the pressure strain correlation.
The ARSM is obtained from~\eqref{eq:rst} by four physical approximations. \emph{Weak equilibrium}~\citep{Rodi1976} assumes the anisotropy is in local equilibrium with the mean strain: it is constant along mean streamlines ($\mathrm{D}a_{ij}/\mathrm{D}t\!\approx\!0$) and its turbulent and molecular diffusive transport is negligible.  This is justified by a separation of timescales: the turbulence relaxation timescale $k/\varepsilon$ is short compared to the timescale on which the mean flow varies along a streamline, so the anisotropy structure equilibrates with the local strain faster than the strain itself changes. Writing $\langle u_i' u_j'\rangle = k(a_{ij} + \tfrac{2}{3}\delta_{ij})$, where $k$ is the turbulent kinetic energy, and substituting the $k$-equation, the left-hand side of~\eqref{eq:rst} reduces to $(\langle u_i' u_j'\rangle/k)(\mathcal{P} - \varepsilon)$, replacing six transport PDEs with algebraic equations. \emph{Isotropic dissipation} sets $\varepsilon_{ij} = \tfrac{2}{3}\varepsilon\delta_{ij}$. The Rotta model~\citep{rotta1951} closes the slow pressure strain as $\Pi_{ij}^{(s)} = -c_1\,\varepsilon\, a_{ij}$, with $c_1$ controlling the rate of return to isotropy. The general linear model of~\citet{launder1975} closes the rapid pressure strain and introduces the coefficient $c_2$. Substituting these into~\eqref{eq:rst} yields the implicit algebraic equation for $a_{ij}$:
\begin{equation}\label{eq:arsm_general}
    N\,\mathbf{a} = -A_1\,\mathbf{S}^{*}
    + \left(\mathbf{a}\boldsymbol{\Omega}^{*}
      - \boldsymbol{\Omega}^{*}\mathbf{a}\right)
    - A_2\!\left(\mathbf{a}\mathbf{S}^{*} + \mathbf{S}^{*}\mathbf{a}
      - \tfrac{2}{3}\,\tr\!\left\{\mathbf{a}\mathbf{S}^{*}\right\}
        \mathbf{I}\right),
\end{equation}
where
$N = A_3 - A_4\,\tr\!\left\{\mathbf{a}\mathbf{S}^{*}\right\}$
and the closure constants are
\begin{equation}\label{eq:closure_constants}
    A_1 = \frac{88}{15(7c_2 + 1)}, \qquad
    A_2 = \frac{5 - 9c_2}{7c_2 + 1}, \qquad
    A_3 = \frac{11(c_1 - 1)}{7c_2 + 1}, \qquad
    A_4 = \frac{11}{7c_2 + 1}.
\end{equation}

The normalised strain and rotation tensors are defined as in the main text,
\begin{equation}\label{eq:SO_app}
    S^{*}_{ij} = \frac{\tau}{2}(U_{i,j} + U_{j,i}), \qquad
    \Omega^{*}_{ij} = \frac{\tau}{2}(U_{i,j} - U_{j,i}), \qquad
    \tau = \frac{1}{\beta^{*}\omega}.
\end{equation}

Equation~\eqref{eq:arsm_general} is a nonlinear implicit tensor
equation that couples the anisotropy $\mathbf{a}$ with the normalised
strain $\mathbf{S}^{*}$ and rotation $\boldsymbol{\Omega}^{*}$ through
the scalar $N$, which itself depends on $\mathbf{a}$. The empirical
constants $c_1$ and $c_2$ enter through $A_1$--$A_4$.

Equation~\eqref{eq:arsm_general} can be solved point-wise by Newton iteration. Alternatively,~\citet{wallin2000explicit} derive an explicit solution by inserting~\eqref{eq:popebasis} into~\eqref{eq:arsm_general}. Given $(\mathbf{S}^{*}, \boldsymbol{\Omega}^{*}, c_1, c_2)$, the anisotropy is then computed in three steps: (i) solve for the scalar $N$ via~\eqref{eq:Nc}--\eqref{eq:N3d}; (ii) evaluate the $\beta$-coefficients~\eqref{eq:beta_3d}; and (iii) apply the eddy-viscosity and anisotropic split in~\eqref{eq:Cmu_eff}--\eqref{eq:aex_split}.
The $\beta^{(n)}$ are rational functions of $N$, the five invariants, and the closure constants:
\begin{align}
    \beta^{(1)} &= -\tfrac{1}{2}A_1 N(30A_2\mathrm{IV} - 21N\mathrm{II}_\Omega - 2A_2^3\mathrm{III}_S + 6N^3 - 3A_2^2\mathrm{II}_S N)/Q, \nonumber \\
    \beta^{(2)} &= -A_1 A_2(6A_2\mathrm{IV} + 12N\mathrm{II}_\Omega + 2A_2^3\mathrm{III}_S - 6N^3 + 3A_2^2\mathrm{II}_S N)/Q, \nonumber \\
    \beta^{(3)} &= -3A_1(2A_2^2\mathrm{III}_S + 3A_2 N\mathrm{II}_S + 6\mathrm{IV})/Q, \nonumber \\
    \beta^{(4)} &= -A_1(2A_2^3\mathrm{III}_S + 3A_2^2 N\mathrm{II}_S + 6A_2\mathrm{IV} - 6N\mathrm{II}_\Omega + 3N^3)/Q, \nonumber \\
    \beta^{(5)} &= 9A_1 A_2 N^2/Q, \quad
    \beta^{(6)} = -9A_1 N^2/Q, \quad
    \beta^{(7)} = 18A_1 A_2 N/Q, \nonumber \\
    \beta^{(8)} &= 9A_1 A_2^2 N/Q, \quad
    \beta^{(9)} = 9A_1 N/Q, \quad
    \beta^{(10)} = 0,
    \label{eq:beta_3d}
\end{align}
with denominator
\begin{align}
    Q &= 3N^5 + \left(-\tfrac{15}{2}\mathrm{II}_\Omega - \tfrac{7}{2}A_2^2\mathrm{II}_S\right)N^3 + (21A_2\mathrm{IV} - A_2^3\mathrm{III}_S)N^2 \nonumber \\
    &\quad + (3\mathrm{II}_\Omega^2 - 8A_2^2\mathrm{II}_S\mathrm{II}_\Omega + 24A_2^2\mathrm{V} + A_2^4\mathrm{II}_S^2)N \nonumber \\
    &\quad + \tfrac{2}{3}A_2^5\mathrm{II}_S\mathrm{III}_S + 2A_2^3\mathrm{IV}\,\mathrm{II}_S - 2A_2^3\mathrm{II}_\Omega\mathrm{III}_S - 6A_2\mathrm{IV}\,\mathrm{II}_\Omega.
    \label{eq:Q_3d}
\end{align}
The scalar $N$ is obtained from the two-dimensional cubic
\begin{equation}\label{eq:Nc}
    N_c =
    \begin{cases}
    \dfrac{A_3}{3} + (P_1 + \sqrt{P_2})^{1/3}
      + \mathrm{sign}(P_1 - \sqrt{P_2})\,
        |P_1 - \sqrt{P_2}|^{1/3},
      & P_2 \geq 0, \\[10pt]
    \dfrac{A_3}{3}
      + 2(P_1^2 - P_2)^{1/6}
        \cos\!\left(\tfrac{1}{3}\arccos\!\left(
          \dfrac{P_1}{\sqrt{P_1^2 - P_2}}\right)\right),
      & P_2 < 0,
    \end{cases}
\end{equation}
with
\begin{align}
    P_1 &= \left(\frac{A_3^2}{27}
      + \left(\frac{A_1 A_4}{6}
        - \frac{2}{9}A_2^2\right)\mathrm{II}_S
      - \frac{2}{3}\,\mathrm{II}_\Omega\right) A_3, \nonumber \\
    P_2 &= P_1^2
      - \left(\frac{A_3^2}{9}
        + \left(\frac{A_1 A_4}{3}
          + \frac{2}{9}A_2^2\right)\mathrm{II}_S
        + \frac{2}{3}\,\mathrm{II}_\Omega\right)^{\!3},
    \label{eq:P1P2}
\end{align}
and a three-dimensional correction
\begin{equation}\label{eq:N3d}
    N = N_c + \frac{162(\phi_1 + \phi_2\, N_c^2)}{D},
    \qquad
    \phi_1 = \mathrm{IV}^2,
    \quad
    \phi_2 = \mathrm{V} - \tfrac{1}{2}\,\mathrm{II}_S\,\mathrm{II}_\Omega,
\end{equation}
\begin{equation}
    D = 20 N_c^4(N_c - \tfrac{1}{2}A_3)
      - \mathrm{II}_\Omega(10 N_c^3 + 15 A_3 N_c^2)
      + 10 A_3\,\mathrm{II}_\Omega^2.
\end{equation}

For numerical stability, the Reynolds stress is split into an effective eddy-viscosity contribution and an explicit anisotropic correction,
\begin{equation}\label{eq:stress_split}
    \langle u_i' u_j'\rangle
    = k\!\left(\tfrac{2}{3}\delta_{ij}
      - 2C_\mu^{\mathrm{eff}} S^{*}_{ij}
      + a^{(\mathrm{ex})}_{ij}\right),
\end{equation}
with
\begin{equation}\label{eq:Cmu_eff}
    C_\mu^{\mathrm{eff}}
      = -\tfrac{1}{2}\!\left(\beta^{(1)}
        + \mathrm{II}_\Omega\,\beta^{(6)}\right),
\end{equation}
absorbed into the turbulent viscosity
\begin{equation}\label{eq:nu_t}
    \nut = C_\mu^{\mathrm{eff}}\,k\,\tau,
\end{equation}
which enters the implicit diffusion operator in~\eqref{eq:rans_main}.
The remaining terms form the explicit anisotropic correction
\begin{equation}\label{eq:aex_split}
    a^{(\mathrm{ex})}_{ij}
      = \sum_{\substack{n=2 \\ n \neq 6}}^{9}
          \beta^{(n)}\,\hat{T}^{(n)}_{ij}
        + \beta^{(6)}\!\left(\mathbf{S}^{*}\boldsymbol{\Omega}^{*2}
          + \boldsymbol{\Omega}^{*2}\mathbf{S}^{*}
          - \tfrac{2}{3}\,\mathrm{IV}\,\mathbf{I}
          - \mathrm{II}_\Omega\,\mathbf{S}^{*}\right),
\end{equation}
where the modified sixth tensor group accounts for the
$\mathrm{II}_\Omega\,\beta^{(6)}$ contribution already absorbed into
$C_\mu^{\mathrm{eff}}$.

The anisotropy depends on the turbulent timescale $\tau = 1/(\beta^{*}\omega)$, which is supplied by the $k$--$\omega$ transport equations~\eqref{eq:komega}. Default coefficients are $\beta^{*} = 0.09$, $\beta_0 = 3/40$, $\gamma = 5/9$, $\sigma_k = \sigma_\omega = 1/2$, $c_1 = 1.8$ and $c_2 = 5/9$.

Every $\beta^{(n)}$ in~\eqref{eq:beta_3d} is a rational function of the invariants and the closure constants. The Jacobian
$\partial \beta^{(n)}/\partial c_{1,2}$ is therefore a nontrivial
function of the invariants, derived directly from the physics, as
anticipated in the opening of this appendix.

\subsection{Neural closure architecture}
\label{app:nn_arch}

The neural network~\citep{dehtyriov2024orans} maps
$\mathbf{z} = (\mathrm{II}_S, \mathrm{II}_\Omega, \mathrm{III}_S, \mathrm{IV}, \mathrm{V}, \log(1{+}\mathrm{Re}_T))$, $\mathrm{Re}_T = k/(\omega\nu)$, to the seven $k$--$\omega$/ARSM coefficients $(\beta^*, \beta_0, \gamma, \sigma_k, \sigma_\omega, c_1, c_2)$ via three $\tanh$ hidden layers of width $H$ interleaved with two input-conditioned $\tanh$ gates:
\begin{equation}\label{eq:gated_mlp}
\begin{aligned}
H_1 &= \tanh(W_1 \mathbf{z} + b_1), \\
H_2 &= \tanh(W_2 H_1 + b_2), \\
\tilde H_2 &= G_1 \odot H_2, & G_1 &= \tanh(W_{G_1}\mathbf{z} + b_{G_1}), \\
H_3 &= \tanh(W_3 \tilde H_2 + b_3), \\
\tilde H_3 &= G_2 \odot H_3, & G_2 &= \tanh(W_{G_2}\mathbf{z} + b_{G_2}), \\
f_\theta &= W_4 \tilde H_3 + b_4.
\end{aligned}
\end{equation}
Inputs are normalised per case to $\mathcal{O}(1)$. Each coefficient is obtained via $p_j = p_j^{\mathrm{ref}}\bigl(1 + C(\sigma(f_{\theta,j}) - \tfrac{1}{2})\bigr)$ with $C = 1.5$, constraining $p_j \in (0.25, 1.75)\,p_j^{\mathrm{ref}}$.

\section{Forward solver: implementation and validation}
\label{app:implementation}

The forward solver is a semi-implicit ADI scheme on a MAC staggered finite-volume grid, implemented in a differentiable framework (PyTorch). We give here the essentials needed to interpret the adjoint derivation of \Cref{app:adjoint}; the discretisation follows standard finite-volume practice~\citep{ferziger2002computational}.

\paragraph{Grids and boundary conditions.}
The square duct is discretised on a Cartesian MAC grid of the full cross-section with $\tanh$-stretching clustering cells toward the four walls. The periodic hills use a body-fitted curvilinear grid, algebraically mapped to a rectangular $(\xi,\eta)$ computational domain; the streamwise direction is periodic, and the implicit tridiagonal solve along that direction is closed by a Sherman--Morrison correction. All walls impose no-slip on velocity, $k \to 0$, and the Menter asymptotic wall value $\omega_w = 6\nu/(\beta_0\Delta y_1^2)$~\citep{wilcox1988reassessment}.

\paragraph{Spatial discretisation.}
Convection uses a second-order van-Leer MUSCL scheme and diffusion is second-order central, with effective viscosity $\nu + \nut$ harmonically averaged at cell faces. The $k$--$\omega$ source terms are treated semi-implicitly: production is evaluated explicitly and destruction is linearised into the diagonal, following standard practice for stiff turbulence transport equations. Streamwise momentum is driven by a uniform body force adapted at each iteration to maintain a target bulk velocity.

\paragraph{Time integration and pressure coupling.}
A single pseudo-time step uses an ADI splitting that advances each field implicitly along one coordinate direction at a time, pressure-projects the in-plane velocities (MAC divergence and symmetric-Laplacian Poisson solve via BiCGSTAB), and then advances $(k, \omega)$ as a coupled $2{\times}2$ block along each direction. This defines the fixed-point map $\G$, iterated to steady state.

\subsection{Validation against reference data}
\label{app:forward_validation}

Before using our forward solver for closure training, we verify that it reproduces established $k$--$\omega$ and EARSM results on both geometries.

\begin{figure}[t]
    \centering
    \includegraphics[width=\textwidth]{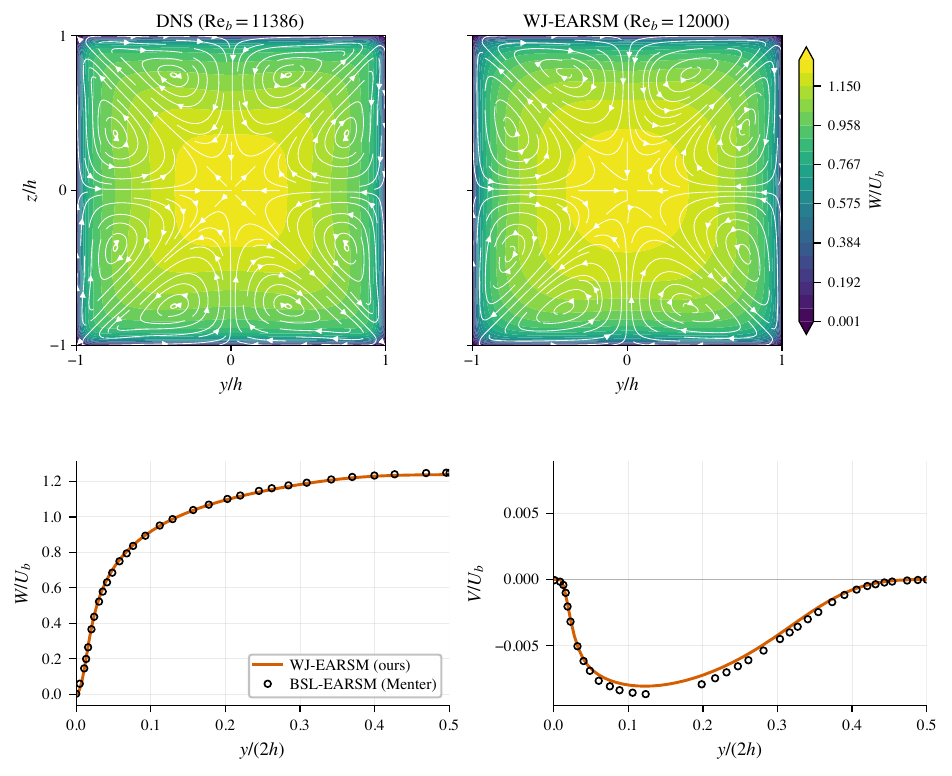}
    \caption{Square-duct forward-solver validation. \textbf{Top}: streamwise velocity $W/U_b$ (colour) with in-plane $(u,v)$ secondary-flow streamlines, showing the characteristic eight-vortex pattern in the DNS \citep{vinuesa2018turbulent} at $\Reb=11386$ (left) and in our WJ-EARSM solution (right). \textbf{Bottom}: profiles of the streamwise $W$ and cross-plane $V$ components along the corner diagonal, compared to BSL-EARSM reference data~\citep{menter2012}. Our solver matches the BSL-EARSM reference to visual accuracy.}
    \label{fig:duct_validation}
\end{figure}

\paragraph{Square duct.}
\Cref{fig:duct_validation} validates the EARSM forward path against an independent reference: Menter's BSL-EARSM implementation, from which we reproduce diagonal-profile reference data for the streamwise $W$ and cross-plane $V$ components at $\Reb=12000$. The top row shows that our WJ-EARSM solver (right) reproduces the eight-vortex secondary-flow pattern seen in the DNS at the nearest matched Reynolds number $\Reb{=}11386$ (left): two counter-rotating vortices per corner driven by the Reynolds-stress anisotropy, which isotropic $k$--$\omega$ cannot generate. The bottom row shows quantitative agreement with the BSL-EARSM reference along the duct diagonal across $y/(2h)\in[0,0.5]$, including the characteristic $V$ undershoot near the wall.

\begin{figure}[t]
    \centering
    \includegraphics[width=\textwidth]{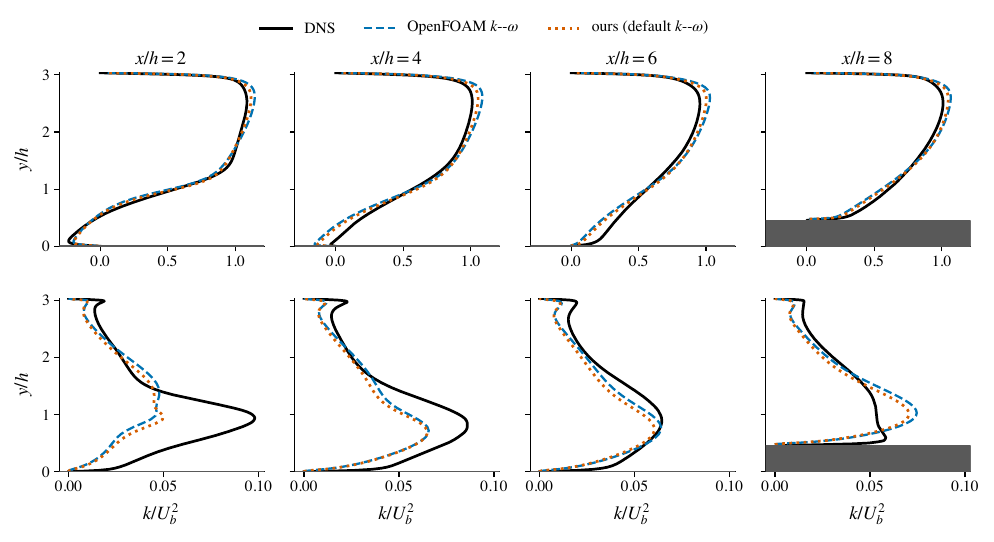}
\caption{Periodic-hills forward-solver validation at $\alpha{=}1.0$, $\Reb{=}5600$.
Streamwise velocity $u/U_b$ (top) and turbulent kinetic energy $k/U_b^2$
(bottom) at stations $x/h\in\{2,4,6,8\}$: DNS~\citep{breuer2009flow} (solid
black), OpenFOAM $k$--$\omega$ (dashed blue) and our PyTorch $k$--$\omega$
(dotted vermilion). The two RANS implementations overlap to within line width,
verifying our curvilinear discretisation.}    \label{fig:hills_validation}
\end{figure}

\paragraph{Periodic hills.}
\Cref{fig:hills_validation} validates the curvilinear-grid $k$--$\omega$ forward path against an independent OpenFOAM implementation of the same closure on the same geometry. Our solver reproduces the OpenFOAM $k$--$\omega$ profiles for both $u$ and $k$ at every station, confirming that the curvilinear discretisation, periodic boundary conditions, and adaptive forcing in our implementation are consistent with a production CFD code. Both RANS solutions miss the target data, which motivates the closure-learning task in \Cref{sec:hills}.

\section{Discrete adjoint of a multistep iterative differentiable solver}
\label{app:adjoint}

This appendix gives the full Lagrangian derivation of the adjoint system summarised in \Cref{sec:training}, the chain-rule decomposition for the hybrid adjoint into per-sub-step pieces (\Cref{fig:adjoint_pipeline}), and the application of that decomposition to the semi-implicit ADI iteration used here.

\subsection{Derivation via the Lagrangian}
\label{app:lagrangian}

From the fixed-point condition, the residual is
$\R(\vU, \vtheta) \coloneqq (\G(\vU;\vtheta) - \vU)/\Delta t$, which vanishes at convergence.
Since $\vU$ satisfies $\R(\vU, \vtheta) = 0$, we introduce the Lagrangian
\begin{equation}
    \Lin(\vU, \vtheta, \vlam) = \J(\vU) - \vlam^\top \R(\vU, \vtheta),
    \label{eq:lagrangian}
\end{equation}
where $\vlam$ is a vector of adjoint variables.
Stationarity with respect to $\vU$ yields the adjoint equation~\eqref{eq:adjoint_system}, a linear system of dimension $n$.
The Jacobian $\partial \R / \partial \vU$ is never formed; \eqref{eq:adjoint_system} is solved iteratively using only matrix-vector products with $(\partial \R / \partial \vU)^\top$ (\Cref{app:adjoint_sweep,app:adjoint_assembly}).
Once $\vlam$ is obtained, differentiating $\Lin$ with respect to $\vtheta$ at $\R = 0$ gives the parameter gradient
$\nabla_{\vtheta} \J = -\vlam^\top (\partial \R / \partial \vtheta)$.

The adjoint computes the same gradient as backpropagation through $K$ unrolled solver steps, but the adjoint formulation collapses the depth-$K$ backpropagation into a single linear solve.
Applying the chain rule through $K$ unrolled steps yields a truncated Neumann series
\begin{equation}
    \nabla_{\vtheta}^{(K)} \J = \dd{\J}{\vU} \sum_{k=0}^{K-1} \left(\dd{\G}{\vU}\right)^{\!k} \dd{\G}{\vtheta},
    \label{eq:neumann}
\end{equation}
which converges to the exact gradient as $K \to \infty$ provided $\rho(\partial \G / \partial \vU) < 1$. Computing \eqref{eq:neumann} by AD through the unrolled iterations stores the intermediate states $\vU^{(0)},\ldots,\vU^{(K)}$, giving $\mathcal{O}(Kn)$ memory; for stiff PDE systems where $K$ is between $10^3$ and $10^5$, this cost is prohibitive.
A second issue arises when the unrolled iteration contains an inner iterative solver, here the BiCGSTAB pressure projection: AD differentiates the truncated inner iteration rather than the converged operator, returning an approximate transpose whose error tracks the inner-solver tolerance (\Cref{sec:scaling}).
The adjoint system \eqref{eq:adjoint_system} avoids both issues.

\subsection{Single-step hybrid adjoint}
\label{app:single_step}

Consider a single forward step
\begin{equation}\label{eq:HG_app}
\mathbf{H}_m(\vU_{m-1},\vtheta)\,\boldsymbol{\delta}_m = \mathbf{g}_m(\vU_{m-1},\vtheta), \qquad \vU_m = \vU_{m-1} + \boldsymbol{\delta}_m,
\end{equation}
with an incoming state-adjoint $\hat{\vU}_m$. The state update $\vU_m = \vU_{m-1} + \boldsymbol{\delta}_m$ has Jacobian $\mathbf{I}$, so $\hat{\boldsymbol{\delta}}_m = \hat{\vU}_m$. Reverse-mode differentiation of the linear solve $\boldsymbol{\delta}_m = \mathbf{H}_m^{-1}\mathbf{g}_m$ then gives
\begin{equation}
    \mathbf{H}_m^\top \hat{\mathbf{g}}_m = \hat{\vU}_m, \qquad \hat{\mathbf{H}}_m = -\hat{\mathbf{g}}_m\,\boldsymbol{\delta}_m^\top.
    \label{eq:adj_sensitivities}
\end{equation}

The first equation is implicit and has the same structure as the forward, so the same efficient numerical solver (Thomas, block-Thomas, or BiCGSTAB) is reused. To propagate the adjoint to $(\vU_{m-1},\vtheta)$, form the scalar
\begin{equation}
    \Psi_m = \hat{\mathbf{g}}_m^\top \mathbf{g}_m(\vU_{m-1},\vtheta) + \hat{\mathbf{H}}_m : \mathbf{H}_m(\vU_{m-1},\vtheta),
    \label{eq:psi_sweep}
\end{equation}
treating $\hat{\mathbf{g}}_m$ and $\hat{\mathbf{H}}_m$ as constants, and differentiate it by reverse-mode AD with respect to $(\vU_{m-1},\vtheta)$. The output combines with the direct contribution from $\vU_m = \vU_{m-1}+\boldsymbol{\delta}_m$ to give
\begin{equation*}
    \hat{\vU}_{m-1} = \hat{\vU}_m + \partial\Psi_m/\partial\vU_{m-1}, \qquad \hat{\vtheta}_m = \partial\Psi_m/\partial\vtheta.
\end{equation*}
The method is hybrid because step \eqref{eq:adj_sensitivities} reuses the forward numerical solver, while \eqref{eq:psi_sweep} is differentiated by reverse-mode AD.

Note that $\mathbf{H}_m$ may itself depend on $\vtheta$; the Lagrangian derivation requires only that the adjoint system \eqref{eq:adjoint_system} is solvable at the converged state, not that $\R$ is separable in $\vU$ and $\vtheta$.

\subsection{Chain-rule adjoint of a composed solver}
\label{app:adjoint_sweep}

When one solver iteration is a composition of $M$ substeps,
\begin{equation}
    \vU^{(k+1)} = \G(\vU^{(k)};\vtheta) = \bigl(\G_M \circ \cdots \circ \G_1\bigr)(\vU^{(k)};\vtheta),
    \qquad \vU_m = \G_m(\vU_{m-1};\vtheta),
    \label{eq:solver_composition}
\end{equation}
each $\G_m$ having the structure of \eqref{eq:HG_app}, the chain rule gives
\begin{equation}
    \Bigl(\dd{\G}{\vU}\Bigr)^{\!\top}
    = \Bigl(\dd{\G_1}{\vU_0}\Bigr)^{\!\top}\Bigl(\dd{\G_2}{\vU_1}\Bigr)^{\!\top}\!\cdots\Bigl(\dd{\G_M}{\vU_{M-1}}\Bigr)^{\!\top},
    \label{eq:adj_chain}
\end{equation}
and similarly for the parameter-Jacobian contribution. The adjoint of one full iteration applies each substep's transpose Jacobian (\Cref{app:single_step}) in reverse order, starting from the incoming state adjoint $\hat{\vU}_M$ and producing $\hat{\vU}_0$ together with partial contributions to the parameter gradient $\{\hat{\vtheta}_m\}_{m=1}^M$. Because each $\mathbf{H}_m$ has structure (tridiagonal for a directional sweep, $2\!\times\!2$ block-tridiagonal for coupled equations and symmetric positive semi-definite for a pressure projection), the cost of the adjoint iteration matches the forward iteration up to a constant.

\subsection{Adjoint solve and gradient extraction}
\label{app:adjoint_assembly}

The chain of \Cref{app:adjoint_sweep} computes the matrix-vector product $\vv\mapsto(\partial\R/\partial\vU)^\top\vv$. The adjoint system \eqref{eq:adjoint_system} is then solved by any matrix-free iterative method that uses this product; we use GMRES~\citep{saad2003iterative}, though iterating the chain to a fixed point also converges to $\vlam$. Each outer iteration sets $\hat{\vU}_M = \vv$, applies the chain in reverse, returning
\begin{equation}
    (\partial\R/\partial\vU)^\top\vv = (\hat{\vU}_0 - \vv)/\Delta t,
\end{equation}
derived from the residual $\R = (\G - \vU)/\Delta t$.

Once $\vlam$ has converged, the chain is applied once more with $\hat{\vU}_M = \vlam$. The sub-step recipe of \Cref{app:single_step} produces parameter contributions $\{\hat{\vtheta}_m\}_{m=1}^M$ that sum to
\begin{equation}
    \nabla_{\vtheta}\J = -\vlam^\top\frac{\partial\R}{\partial\vtheta} = -\frac{1}{\Delta t}\sum_{m=1}^M \hat{\vtheta}_m.
    \label{eq:gradient_per_link}
\end{equation}

\Cref{fig:adjoint_pipeline} in the main text schematically illustrates the hybrid adjoint.

\paragraph{Closure factoring.}\label{app:earsm_factoring}

Let $f_{\vtheta}$ denote the embedded neural network used in the assembly of $(\mathbf{H}_m, \mathbf{g}_m)$ (\Cref{fig:adjoint_pipeline}). The recipe of \Cref{app:single_step} differentiates $\Psi_m$ against $\vtheta$ at every sub-step, which traces through $f_{\vtheta}$ at every one of the $M$ sub-steps per gradient computation. In practice the implementation factors $f_{\vtheta}$ out of the chain:

\begin{enumerate}[nosep]
    \item Evaluate the network outputs $\mathbf{c} = f_{\vtheta} (\vU)$ once at the converged state and detach $\mathbf{c}$ from the AD graph (treat it as a constant input to the assembly).
    \item In each sub-step of the chain, differentiate $\Psi_m$ against $(\vU_{m-1}, \mathbf{c})$ rather than $(\vU_{m-1}, \vtheta)$. The sub-step returns the state-adjoint $\hat{\vU}_{m-1}$ as before, and a partial adjoint at the closure outputs $\hat{\mathbf{c}}_m \coloneqq \partial\Psi_m/\partial\mathbf{c}$. The network sits outside this AD graph, so the per sub-step cost is independent of network depth.
    \item Accumulate $\hat{\mathbf{c}} = \sum_{m=1}^M \hat{\mathbf{c}}_m$ across the chain.
    \item Form the scalar $\Psi=-\frac{1}{\Delta t}\hat{\mathbf{c}}^\top f_{\vtheta} (\vU)$, treating $\hat{\mathbf{c}}$ as a constant, and backpropagate $\Psi$; this is one reverse-mode pass through $f_{\vtheta}$ to obtain  $\nabla_{\vtheta}\J$.
\end{enumerate}

The result equals $-\frac{1}{\Delta t}\sum_m \hat{\vtheta}_m$ from \eqref{eq:gradient_per_link}: the chain rule $\partial\Psi_m/\partial\vtheta = (\partial\Psi_m/\partial\mathbf{c})(\partial\mathbf{c}/\partial\vtheta)$ pulls $\partial\mathbf{c}/\partial\vtheta$ out of the per-sub-step sum (the network Jacobian is the same at every sub-step), leaving $\hat{\mathbf{c}}$ behind. The total network cost is one backpropagation per gradient computation rather than $M$. The saving is small for the networks used here, but it makes the factoring applicable to larger closures.

\subsection{Differentiable RANS Application: semi-implicit ADI with pressure projection}
\label{app:adi_application}
\paragraph{Time integration and pressure coupling.}
A single pseudo-time step uses an ADI splitting that advances each field implicitly along one coordinate direction at a time. With $\Delta\phi = \phi^{*}-\phi^{n}$ the sub-iterate increment, each half-sweep solves the delta-form system
\begin{equation}
  \biggl[\frac{\mathbf{I}}{\Delta t}
        + \delta_\xi\!\Bigl(\tfrac{\partial f_c}{\partial\phi}
                          - \tfrac{\partial f_v}{\partial\phi}\Bigr)^{\!n}
        - \Bigl(\tfrac{\partial s}{\partial\phi}\Bigr)^{\!n}\biggr]\Delta\phi
  = -\delta_\xi\!\bigl(f_c-f_v\bigr)^{n}
    -\delta_\eta\!\bigl(f_c-f_v\bigr)^{n}
    + s(\phi^{n}),
  \label{eq:adi_delta_form}
\end{equation}
where $\xi$ is the sweep direction, $\eta$ the cross-direction (lagged at the latest iterate); $\delta_\xi,\delta_\eta$ are discrete finite-volume divergences on the staggered MAC faces; $f_c,f_v$ are the convective and viscous fluxes; $s$ collects sources (turbulent production for $k,\omega$ and the EARSM anisotropic-stress contribution); superscript $n$ denotes the current sub-iterate.
For momentum we retain only the viscous Jacobian on the left, giving a scalar tridiagonal solve with convection fully explicit.
For the coupled $(k,\omega)$ block we additionally retain the first-order upwind convective Jacobian (with the second-order correction deferred to the right-hand side) and linearise the destruction terms pointwise into the $2\!\times\!2$ block diagonal, giving a block-tridiagonal solve.
After the in-plane momentum sweeps a fractional-step MAC projection $\Pi$ enforces $\nabla\!\cdot\!\mathbf{u}=0$: a symmetric Poisson solve (BiCGSTAB) on the diagonal Laplacian, wrapped on the curvilinear grid by a non-orthogonal corrector loop on the cross-Laplacian.

The forward iteration fits \eqref{eq:solver_composition} with $M = 9$:
\begin{equation}
  \G = \G_{k\omega,y} \circ \G_{k\omega,x} \circ \G_{w,y} \circ \G_{w,x} \circ \Pi \circ \G_{v,y} \circ \G_{v,x} \circ \G_{u,y} \circ \G_{u,x},
  \label{eq:adi_composition}
\end{equation}
where each $\G_{\phi,\xi}$ is an ADI half-sweep advancing a single field $\phi\in\{u,v,w\}$ (tridiagonal $\mathbf{H}$) or the coupled pair $(k,\omega)$ ($2\!\times\!2$ block-tridiagonal $\mathbf{H}$) implicitly along direction $\xi$, and $\Pi$ is the MAC pressure projection. Each half-sweep assembles $(\mathbf{H}_m,\mathbf{g}_m)$ from convection, diffusion, source, metric, boundary, and EARSM contributions through a differentiable assembly routine. The composition \eqref{eq:adi_composition} defines the fixed-point map $\G$.

\paragraph{Adjoint half-sweep.}
The adjoint half-sweep inherits the structure of \eqref{eq:adi_delta_form} by transposition:
\begin{equation}
  \biggl[\frac{\mathbf{I}}{\Delta t}
        + \delta_\xi\!\Bigl(\tfrac{\partial f_c}{\partial\phi}
                          - \tfrac{\partial f_v}{\partial\phi}\Bigr)^{\!\top}
        - \Bigl(\tfrac{\partial s}{\partial\phi}\Bigr)^{\!\top}\biggr]
        \hat{\mathbf{g}} = \hat{\vU}_m,
  \label{eq:adi_delta_form_adj}
\end{equation}
solved at the same cost as the forward sweep (transpose-Thomas for momentum, transpose block-Thomas for $(k,\omega)$, with a Sherman--Morrison transpose correction in the periodic streamwise direction).

The pressure projection $\Pi$ chains MAC divergence, a symmetric-Laplacian Poisson solve, a MAC gradient, and a mean-subtraction. Its implicit operator is the diagonal symmetric Laplacian, $L_\textrm{diag}^\top = L_\textrm{diag}$, so the same BiCGSTAB solver is reused for the transpose. On the curvilinear grid the non-orthogonal corrector iteration is applied in reverse explicitly, with each application of the cross-Laplacian transposed by AD (treated like any other assembly). The EARSM block is a pure algebraic function $(\vU,\vtheta)\mapsto(\nu_t,\,a^{(\mathrm{ex})}_{ij},\,\text{source terms})$ containing the Wallin--Johansson cubic and the embedded neural network. It appears in \eqref{eq:adi_composition} as differentiable operations within the assembly routines, so its transpose is captured entirely by the AD pass of \eqref{eq:psi_sweep}. No manual derivation is required for the cubic, the $\beta$-coefficients, or the network.

\subsection{Adjoint validation}
\label{app:vjp}

\begin{figure}[t]
    \centering
    \includegraphics[width=0.55\textwidth]{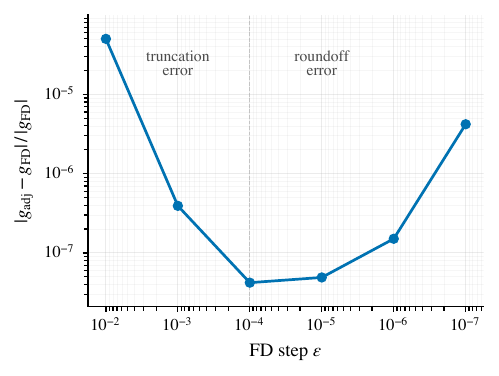}
\caption{End-to-end finite-difference verification of the hybrid adjoint
gradient. Pointwise relative error
$|g_\mathrm{adj} - g_\mathrm{FD}|/|g_\mathrm{FD}|$ for
$\partial \J / \partial c_1$ versus FD step $\varepsilon$ on a $32^2$
square-duct grid (forward/adjoint tolerances $10^{-12}$). Left branch:
truncation-limited ($\propto \varepsilon^{2}$). Right branch:
roundoff-limited ($\propto 1/\varepsilon$). Minimum relative error
$4.2\times 10^{-8}$ at $\varepsilon{=}10^{-4}$; the plateau floor is
consistent with the adjoint GMRES residual.}
\label{fig:vcurve}
\end{figure}

We verify each adjoint building block against central finite differences using
the vector--Jacobian identity
$\langle \hat{\vu}_{\mathrm{out}},\, \text{d}\vu_{\mathrm{out}}\rangle
  = \langle \hat{\vu}_{\mathrm{in}},\, \text{d}\vu_{\mathrm{in}}\rangle$,
where $\text{d}\vu_{\mathrm{out}}$ is obtained by central differences along a
random perturbation of $\vu_{\mathrm{in}}$, sweeping the FD step
$\epsilon \in [10^{-2}, 10^{-10}]$ and reporting the V-curve minimum.
Operations tested: each momentum half-sweep ($u,v,w$ in both $x$ and $y$), the
coupled $k$--$\omega$ block sweeps, the pressure projection, the EARSM block,
the adaptive forcing, and the composed forward step. Each passes at
$<10^{-5}$ relative error in double precision. The end-to-end gradient check
is in \Cref{fig:vcurve}.

\subsection{Training procedure}
\label{app:training_procedure}

With the forward solver and adjoint verified, the outer training loop proceeds as follows.
The adjoint system is solved by GMRES to relative residual $10^{-6}$ (capped at 2000 iterations), reusing the symmetric pressure Laplacian in the transpose projection. Parameters are updated by full-batch BFGS~\citep{nocedal2006numerical} using the exact adjoint gradient; gradient clipping is applied at norm $1.0$. Before the adjoint loop begins, the network is pretrained by $5{,}000$ supervised steps to reproduce the default EARSM coefficients, which places $\vtheta$ in a stable basin at the start of training. An adjoint solve costs a quarter to a third of a forward solve. At deployment the network adds about $2\%$ to the cost of each solver iteration.

\section{Second PDE: convection--diffusion--reaction}
\label{app:cdr}

To test that the recipe of \Cref{sec:method,app:adjoint} is genuinely solver-agnostic, we apply it to a PDE structurally unrelated to RANS. Consider a scalar convection--diffusion--reaction equation on the unit square,
\begin{equation}
    \dd{u}{t} + \mathbf{v}\cdot\nabla u
    = D\,\nabla^2 u + R(u;\vtheta) + S(\mathbf{x}),
    \qquad u=0 \text{ on } \partial\Omega,
    \label{eq:cdr}
\end{equation}
with a divergence-free cellular flow $\mathbf{v} = (\sin\pi x\cos\pi y,\,-\cos\pi x\sin\pi y)$, a fixed spatial source $S(\mathbf{x}) = \sin(\pi x)\sin(\pi y)$, and an unknown nonlinear reaction term $R(u;\vtheta)$ approximated by a small MLP. The reference data come from a solver with a bistable reaction $R_\text{true}(u) = u(1-u)(u-0.3)$, evaluated at three diffusivities $D\in\{0.005, 0.01, 0.02\}$ on a $64{\times}64$ grid.

The solver uses the same semi-implicit ADI splitting as the RANS code: at each pseudo-time step, the implicit operator $\mathbf{H}$ is a pair of tridiagonal linear systems (one per coordinate direction) for the diffusion term, and the explicit right-hand side $\mathbf{g}$ carries the convection, reaction, and source. The adjoint is obtained by the same recipe (hand-coded transpose-Thomas for each sub-sweep, autodiff of the assembly of $\mathbf{g}$ and of the neural reaction term) with no methodological modification. Implementation consists of swapping out the forward-residual assembly of the RANS code for the CDR residual; all adjoint machinery is reused unchanged.

We train on a single case ($D=0.01$) and evaluate on two out of sample cases ($D=0.005$ and $D=0.02$), using the same full-batch BFGS loop as the RANS experiments (\Cref{app:training_procedure}) with an area-averaged squared-error loss on $u$. \Cref{tab:cdr} reports training dynamics. The loss decreases by nearly three orders of magnitude on the training case and transfers to both out of sample cases without retraining, confirming that the recipe applies unchanged outside the RANS setting. We report values at epoch $172$, before convergence; both losses were still decreasing.

\begin{table}[h]
\centering
\caption{Single-case training on 2D convection--diffusion--reaction ($64{\times}64$): area-averaged squared-error objective on the training diffusivity $D{=}0.01$ and mean over two out-of-sample diffusivities $D\in\{0.005, 0.02\}$.}
\label{tab:cdr}
\small
\begin{tabular}{lcc}
\toprule
                                        & $\J^\text{train}$ & $\J^\text{test}$ \\
\midrule
Initial                                 & $9.55$     & $18.3$       \\
After training                          & $\mathbf{0.0114}$ & $\mathbf{0.078}$ \\
\midrule
Reduction factor                        & $838\times$ & $235\times$ \\
\bottomrule
\end{tabular}
\end{table}

\section{Ablations and additional diagnostics}
\label{app:ablations}

This section reports DARSM-internal ablations: training dynamics, network-capacity sweep, optimiser comparison, learned closure coefficients, the corrections shared between the duct- and hills-trained networks, and a global-versus-spatial parameterisation ablation.

\subsection{Training dynamics and evaluation}
\label{app:training}

\begin{figure}[t]
    \centering
    \includegraphics[width=0.48\textwidth]{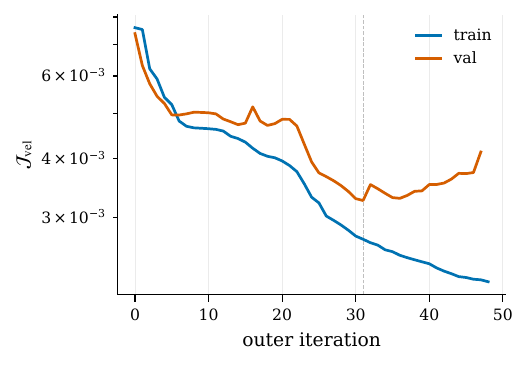}\hfill
    \includegraphics[width=0.48\textwidth]{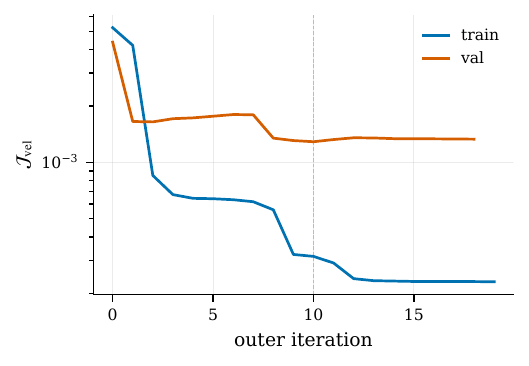}
    \caption{Velocity-only loss $\J_\text{vel}$ over outer iterations for one fold. \textbf{Left}: square duct (Reynolds split, $H{=}10$, BFGS). \textbf{Right}: periodic hills (shape-generalisation split, $H{=}10$, BFGS). Vertical dashed line marks the best-val epoch. Train decreases monotonically; val minimises then drifts.}
    \label{fig:jcurves}
\end{figure}

\Cref{fig:jcurves} shows the velocity-only loss $\J_\text{vel}$ during
training for the square duct (Reynolds split) and periodic hills
(shape-generalisation split). Training loss decreases monotonically in
both cases while validation loss reaches a minimum before overfitting. We address this with leave-one-out cross-validation (LOOCV) over the out of sample cases. In each fold, one out of sample case is designated as the test case and the remaining out of sample cases form the validation set; the model checkpoint that minimises the mean validation loss on this fold is used to evaluate the corresponding test case. We rotate over all folds and report the average out-of-sample velocity error.

\subsection{Network capacity}
\label{app:hsweep}

We assess sensitivity to network capacity by sweeping the hidden width $H \in \{3, 7, 10, 15, 20, 40\}$, retraining from scratch at each value with matched training data, initialisation, and optimisation protocol. Each model is evaluated under the same LOOCV early-stopping protocol as the headline runs (\Cref{app:training}). The width $H$ and the stopping epoch are the only hyperparameters tuned, and both are selected using validation data alone. Every tested $H$ improves on the baseline RANS, with the results insensitive to $H\geq10$ indicating that DARSM is robust to network capacity. We report headline results at $H=10$, the smallest network in the insensitive range.

\subsection{Optimiser: BFGS vs.\ Adam}
\label{app:bfgs_vs_adam}

\begin{figure}[t]
    \centering
    \includegraphics[width=0.55\textwidth]{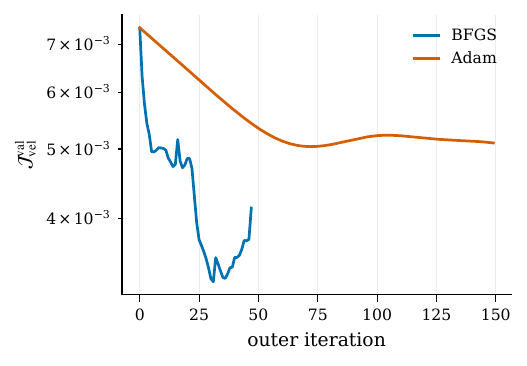}
    \caption{Validation loss trajectory for BFGS (used throughout) vs.\ Adam, matched architecture and initialisation. Exact adjoint gradients are deterministic and full-batch, which favours classical second-order optimisation over stochastic first-order methods.}
    \label{fig:bfgs_vs_adam}
\end{figure}

Since the adjoint returns exact deterministic gradients at every outer iteration, the training loop is effectively full-batch, and classical second-order methods are well-matched to this setting. \Cref{fig:bfgs_vs_adam} compares $\mathcal{J}^\text{val}$ curves for BFGS and Adam~\citep{kingma2015adam} on the same network: BFGS converges in far fewer outer iterations. We use BFGS throughout. Although the adjoint gradients are deterministic, training is not, because seed-to-seed variability enters entirely through the random network initialisation. We therefore train each configuration across $3$ random seeds and report the seed mean out-of-sample error together with the standard deviation.

\subsection{Learned closure coefficients}
\label{app:learned_coeffs}

\begin{figure}[t]
    \centering
    \includegraphics[width=\textwidth]{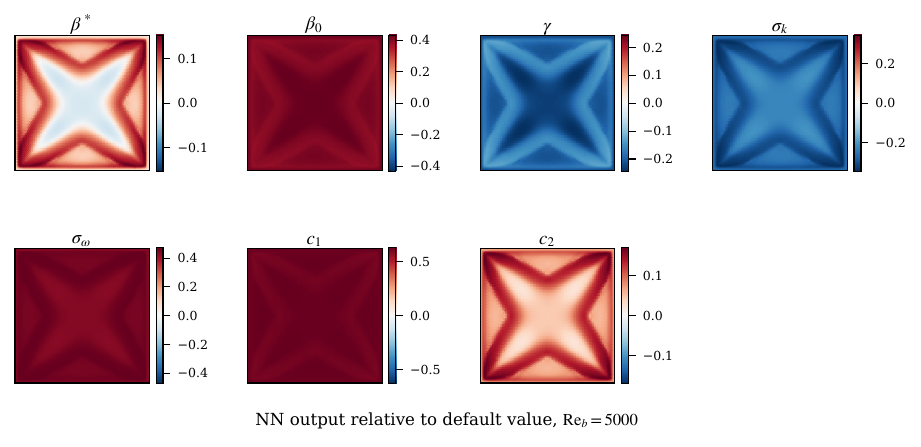}
    \caption{Learned closure corrections at $\Reb{=}5000$: NN output for each of the seven $k$--$\omega$/EARSM coefficients, expressed as fractional deviation from the default value. Red is pushed above default, blue pushed below and white is default. The closure is spatially varying and concentrates its corrections near walls and along corner bisectors, exactly where secondary flow is generated.}
    \label{fig:learned_coeffs_duct}
\end{figure}

\begin{figure}[t]
    \centering
    \includegraphics[width=\textwidth]{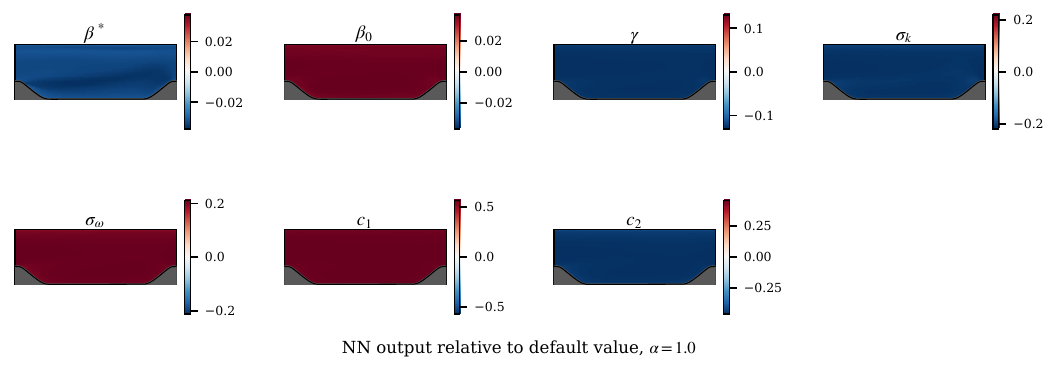}
    \caption{Learned closure corrections on periodic hills at $\alpha{=}1.0$: same quantities as \Cref{fig:learned_coeffs_duct} but on the curvilinear hill geometry. Hill surface is masked in grey. The trained closure alters coefficients most strongly inside the separation bubble and along the reattachment line.}
    \label{fig:learned_coeffs_hills}
\end{figure}

The network outputs seven spatially varying multiplicative corrections to the default EARSM coefficients. \Cref{fig:learned_coeffs_duct,fig:learned_coeffs_hills} show these corrections (as fractional deviation from default) at one representative case per geometry. The largest corrections are nearly uniform increases of $c_1$, $\sigma_\omega$ and $\beta_0$ and decreases of $\gamma$ and $\sigma_k$. The spatial variation is carried mainly by $\beta^*$ and $c_2$ and is concentrated in the near-wall and near-corner regions where the default closure most poorly predicts the target data.

\subsection{Shared and geometry-specific coefficient corrections}
\label{app:coeff_corrections}

To find which learned corrections transfer from the duct to the hills, we evaluate the duct-trained and the hills-trained networks on the same converged flow states for all five hill geometries and report each network's mean correction to each constant (\Cref{tab:coeff_corrections}). For $\gamma$, $\sigma_k$, $\sigma_\omega$ and $c_1$ the two independently trained networks find corrections of the same sign and comparable size, although they were trained on different geometries, so these shared corrections reflect the closure rather than the training flow. For $c_2$ and $\beta_0$ the networks disagree, the hills-trained network reducing $c_2$ and leaving $\beta_0$ unchanged while the duct-trained network leaves $c_2$ nearly unchanged and raises $\beta_0$, so these corrections reflect the training geometry. Both leave $\beta^*$ nearly unchanged.

\begin{table}[h]
\centering
\caption{Mean correction (change from the default value) to each closure constant over the five hill geometries, for the network trained on the duct and the network trained on the hills ($\alpha{=}1.0$); defaults in parentheses.}
\label{tab:coeff_corrections}
\small
\begin{tabular}{lrr}
\toprule
Constant (default) & Duct-trained & Hills-trained \\
\midrule
$\gamma$ ($0.56$)        & $-0.12$ & $-0.10$ \\
$\sigma_k$ ($0.50$)      & $-0.12$ & $-0.12$ \\
$\sigma_\omega$ ($0.50$) & $+0.22$ & $+0.15$ \\
$c_1$ ($1.80$)           & $+1.12$ & $+0.62$ \\
$c_2$ ($0.56$)           & $+0.03$ & $-0.31$ \\
$\beta_0$ ($0.075$)      & $+0.03$ & $+0.00$ \\
$\beta^*$ ($0.09$)       & $+0.00$ & $+0.00$ \\
\bottomrule
\end{tabular}
\end{table}

The mean corrections are nearly uniform over the hills because the invariants the network takes as input are nearly uniform there; where the invariants vary, the corrections vary with them (in the separated shear layer the hills-trained corrections are strongly correlated with the local invariant deviations, and on the duct the same duct-trained network produces the strongly varying fields of \Cref{fig:learned_coeffs_duct}). To test whether the transfer is a broad recalibration, we deployed an EARSM with constants fixed at the duct-trained network's mean corrections on all five geometries. This fixed model is two to three times worse than the duct-trained network on every geometry and at $\alpha{=}0.5$ is worse than the default closure. The shared corrections therefore explain part of the improvement, and the dependence of the corrections on the local flow state explains much of the rest.

\subsection{Spatial irreducibility}\label{app:duct_ablation}

To check whether spatial variation of the closure coefficients is actually necessary, we replace the neural network with a set of global scalar parameters that we optimise with our adjoint method. \Cref{tab:duct_ablation} shows that globally tuning all seven $k$--$\omega$ and EARSM constants recovers only part of the improvement, leaving ${\sim}40\%$ of the error gap versus the full spatial network. The closure correction required by this system is irreducibly spatial.  

\begin{table}[t]
\centering
\caption{Architecture ablation on the Reynolds-number split: $\J^\text{test}
\times 10^{3}$ as closure capacity grows from analytic coefficients to a
neural network. All rows share the same adjoint training loop; only the
closure parameterisation differs. DARSM values are mean$\pm$std over seeds.}
\label{tab:duct_ablation}
\small
\begin{tabular}{lcc}
\toprule
Learnable parameterisation & \# params & $\J^\text{test}$ \\
\midrule
Default EARSM (no training)                & $0$ & $7.38$ \\
Global $(c_1,c_2)$ only (2 scalars)                & $2$ & $6.54$ \\
Global coefficients (7 scalars)                    & $7$ & $5.11$ \\
\textbf{DARSM}                     & $507$ & $\mathbf{3.56\pm0.27}$ \\
\bottomrule
\end{tabular}
\end{table}

\section{Extended empirical results}\label{app:extended_results}

This section reports the application-level metrics summarised in \Cref{sec:duct,sec:hills}.

\subsection{Application-level metrics}
\label{app:app_metrics}

\paragraph{Reynolds-stress anisotropy.}
On the square duct the DNS provides the full stress tensor. \Cref{tab:anisotropy} reports the relative $L_2$ error of each anisotropy component for the default EARSM and DARSM at the six Reynolds numbers. The shear components, which act on the mean flow, improve at every Reynolds number; the normal-stress components are not constrained by the velocity objective and degrade slightly.

\begin{table}[h]
\centering
\caption{Square duct: relative $L_2$ error of the Reynolds-stress anisotropy components against the DNS for the default EARSM and for DARSM.}
\label{tab:anisotropy}
\small
\setlength{\tabcolsep}{4pt}
\begin{tabular}{llcccccc}
\toprule
 & & $\Reb{=}4410$ & $5000$ & $7000$ & $11386$ & $17800$ & $40000$ \\
\midrule
\multirow{2}{*}{$a_{13},a_{23}$ (streamwise shear)} & EARSM & $0.85$ & $0.78$ & $0.64$ & $0.54$ & $0.47$ & $0.44$ \\
 & DARSM & $\mathbf{0.67}$ & $\mathbf{0.61}$ & $\mathbf{0.50}$ & $\mathbf{0.41}$ & $\mathbf{0.34}$ & $\mathbf{0.29}$ \\
\midrule
\multirow{2}{*}{$a_{12}$ (in-plane shear)} & EARSM & $2.16$ & $2.06$ & $1.72$ & $1.66$ & $1.53$ & $1.91$ \\
 & DARSM & $\mathbf{1.45}$ & $\mathbf{1.34}$ & $\mathbf{1.12}$ & $\mathbf{1.11}$ & $\mathbf{1.09}$ & $\mathbf{1.26}$ \\
\midrule
\multirow{2}{*}{$a_{11},a_{22}$ (normal stresses)} & EARSM & $\mathbf{0.81}$ & $\mathbf{0.81}$ & $\mathbf{0.78}$ & $\mathbf{0.74}$ & $\mathbf{0.70}$ & $\mathbf{0.62}$ \\
 & DARSM & $0.87$ & $0.86$ & $0.84$ & $0.82$ & $0.78$ & $0.72$ \\
\bottomrule
\end{tabular}
\end{table}

\paragraph{Separation, reattachment and turbulent kinetic energy on the hills.}
\Cref{tab:hills_metrics} reports the separation onset, the main-bubble reattachment and the $k$ error summarised in \Cref{sec:hills}, computed with the same procedure for the DNS, the default EARSM and DARSM: separation and reattachment from the sign changes of the wall shear stress, evaluated from the wall-tangential first-cell velocity, and the $k$ error as the area-weighted relative $L_2$ norm against the DNS.

\begin{table}[h]
\centering
\caption{Separation onset $x_\mathrm{sep}$ and main-bubble reattachment $x_\mathrm{reatt}$ (hill heights from the crest) and area-weighted relative $L_2$ error of $k$ against the DNS for periodic hills. DARSM is trained on $\alpha{=}1.0$; its values are mean$\pm$std over seven seeds.}
\label{tab:hills_metrics}
\small
\setlength{\tabcolsep}{4pt}
\begin{tabular}{lccc|ccc|cc}
\toprule
 & \multicolumn{3}{c|}{$x_\mathrm{sep}$} & \multicolumn{3}{c|}{$x_\mathrm{reatt}$} & \multicolumn{2}{c}{$k$ error} \\
$\alpha$ & DNS & EARSM & DARSM & DNS & EARSM & DARSM & EARSM & DARSM \\
\midrule
$0.5$   & $0.03$ & $0.13$ & $0.03\pm0.00$ & $6.28$ & $6.58$ & $5.74\pm0.80$ & $0.35$ & $\mathbf{0.29}$ \\
$0.75$  & $0.11$ & $0.19$ & $0.08\pm0.06$ & $5.56$ & $7.13$ & $4.84\pm0.11$ & $0.41$ & $\mathbf{0.26}$ \\
$1.0$ & $0.17$ & $0.25$ & $0.22\pm0.01$ & $4.73$ & $7.61$ & $4.85\pm0.07$ & $0.50$ & $\mathbf{0.28}$ \\
$1.25$  & $0.31$ & $0.33$ & $0.31\pm0.02$ & $4.41$ & $7.95$ & $4.84\pm0.07$ & $0.52$ & $\mathbf{0.30}$ \\
$1.5$   & $0.46$ & $0.42$ & $0.42\pm0.01$ & $4.29$ & $7.28$ & $4.82\pm0.09$ & $0.53$ & $\mathbf{0.32}$ \\
\bottomrule
\end{tabular}
\end{table}

\paragraph{Solver convergence.}
Every DARSM deployment behind \Cref{tab:duct,tab:duct_main,tab:hills} converged, and from a cold start reaches the same $10^{-5}$ tolerance with unchanged solver settings. The network only sets the coefficients of the closure inside the default solver's unchanged implicit treatment, so the stress keeps the form of a closure derived from the Reynolds stress transport equations.

\section{Baseline methods and scaling benchmark}
\label{app:baselines}

This section documents the architectures, hyperparameters, and training protocols for the comparison methods reported in \Cref{tab:duct} and the scaling routes in \Cref{tab:scaling}. All baselines were implemented in double precision and run on the same target data as the adjoint closure.

\subsection{TBNN (tensor-basis neural network) and a-priori NN-EARSM}
\label{app:baseline_apriori}

\paragraph{TBNN.}
We use the architecture of \citet{ling2016reynolds} (8 hidden fully-connected layers, 30 neurons per layer and leaky ReLU), which predicts the coefficients of the ten-term integrity basis of \citet{pope1975more} from the five scalar invariants, and $a_{ij}$ is reconstructed as a linear combination of the ten basis tensors.

\paragraph{A-priori NN-EARSM.}
Same gated 4-layer architecture as our adjoint closure ($H{=}10$), supervised
offline to predict the two EARSM shape coefficients $(c_1, c_2)$; $a_{ij}$ is
then reconstructed from the \citet{wallin2000explicit} algebraic closure rather than
the full Pope basis. The five Wilcox $k$-$\omega$ coefficients are held at
their defaults.

\paragraph{Training.}
Adam with learning rate $10^{-4}$ for $10^{4}$ epochs, full batch, minimising
$\|a_{ij}^\mathrm{NN}(\eta^\mathrm{DNS}) - a_{ij}^\mathrm{DNS}\|^{2}$ summed
over cells and training cases, where $\eta^\mathrm{DNS}$ are invariants
computed from the DNS state.

\paragraph{Deployment.}
At test time the trained model replaces the default EARSM anisotropy in the
forward solver. In the \emph{iterative} case the network is queried at
every forward iteration, but the solver diverges; this is an error-amplification mechanism formally identified by~\citet{wu2019rans} as a consequence of the ill-conditioning of the RANS equations with respect to explicitly-treated Reynolds stress. We therefore report the frozen-injection protocol of~\citet{ling2016reynolds}, in which the network is queried once on the converged default $k$-$\omega$ state and the resulting $a_{ij}$ is held fixed as the solver re-converges.

\subsection{FIML (field inversion machine learning)}
\label{app:baseline_fiml}

We follow the two-step field inversion and machine learning procedure of~\citet{parish2016paradigm}. \textbf{Step~1} (field inversion): per training case, the Wilcox-$k$-$\omega$ destruction coefficient $\beta_0$ (default $3/40$) is replaced by a free scalar field $\beta_0(x,y)$ over the $n_x{\times}n_y$ grid. This field is optimised by the same steady-state adjoint used by our method to minimise $\J_\text{vel}$, with a Tikhonov penalty  $\mu\|\beta_0-3/40\|^{2}$ ($\mu{=}0.1$); the other six closure coefficients are fixed. \textbf{Step~2} (supervised ML): a neural network with the same gated-MLP architecture as the adjoint closure (\Cref{app:nn_arch}) is trained to predict $\beta_0$ from six local flow invariants evaluated at the Step-1-converged RANS state, using Adam with learning rate $10^{-3}$ for $5{\times}10^{4}$ epochs. At deployment, the trained NN replaces $\beta_0$ in the forward solver and is evaluated on all cases.

\subsection{DeepONet (operator network)}
\label{app:baseline_deeponet}

\paragraph{Architecture.}
Branch--trunk DeepONet~\citep{lu2021learning} with one branch/trunk pair per velocity field ($W$, $U$, $V$, each on its own staggered grid). Each branch is an MLP $\mathbb{R}\to\mathbb{R}^{p}$ mapping normalised Reynolds number to a $p$-dimensional coefficient vector; each trunk is an MLP $\mathbb{R}^2\to\mathbb{R}^{p}$ mapping normalised grid coordinates to a $p$-dimensional basis; the field at a point is their inner product plus a learnable scalar bias. Both branch and trunk use two hidden layers of width $H{=}128$ with $\tanh$ activations. We sweep $p\in\{32, 64, 128\}$ ($1.3$--$2.0{\times}10^{5}$ parameters; $\sim 250$--$400\times$ the adjoint closure's $\sim 500$).

\paragraph{Training.}
Adam with constant learning rate $10^{-3}$ for $3{\times}10^{5}$ epochs, full batch. The loss is the adjoint's $\J_\text{vel}$~\eqref{eq:Jvel}, so DeepONet optimises the same scalar reported in \Cref{tab:duct}. Reynolds number and $(x,y)$ are normalised to $[0,1]$ using the full six-case range. We use the training/validation split of \Cref{sec:comparison}, track the best-validation checkpoint (early stopping on $\J_\text{vel}^\text{val}$), and report the best configuration across the $p$-sweep.

\subsection{PINN (physics-informed neural network)}
\label{app:baseline_pinn}

\paragraph{Architecture.}
MLP $\mathbb{R}^{3}\to\mathbb{R}^{6}$ mapping $(\xi, \eta, \mathrm{Re}_\text{norm})$ to $(W, U, V, k, \omega, p)$, with four hidden layers of width $H{=}128$ and $\tanh$ activations ($\approx 5.1{\times}10^{4}$ parameters). Softplus is applied to $k$ and $\omega$ for positivity. $(\xi,\eta)\in[0,1]^{2}$ and Re are normalised using the full six-case range; the network is trained on the two training Reynolds numbers of \Cref{sec:comparison} simultaneously and evaluated at the four out-of-sample cases.

\paragraph{Losses.}
The physics residual enforces the six incompressible RANS $k$--$\omega$ EARSM equations (continuity, three momentum components and $k$- and $\omega$-transport) with \citep{wilcox1988reassessment} coefficients and the EARSM anisotropic stress entering both the momentum divergence and the $k$-production term \citep{wallin2000explicit}. The boundary-condition loss enforces no-slip, $k{\to}0$, and the solver's near-wall $\omega_w$ at $250$ points per wall. The streamwise $\partial P/\partial z$ is computed once from the default EARSM reference via a wall-shear-stress balance and held fixed. We augment the physics and boundary condition loss with a weighted target data loss, which we sweep over $\{0, 0.1, 1, 10\}$, and we report the best-performing setting and training epoch.

\paragraph{Training.}
Two-stage: Adam with learning rate $10^{-3}$ for up to $3{\times}10^{5}$ epochs, then L-BFGS with strong-Wolfe line search for up to $5{\times}10^{3}$ outer steps. Adam-stage collocation uses $5{\times}10^{3}$ points re-sampled uniformly in $[0,1]^{2}\times\{\mathrm{Re}_\text{norm}\}$ each epoch.

\subsection{Source-term neural closure}
\label{app:baseline_source_term}

\paragraph{Architecture.}
A gated 4-layer MLP (the same architecture as the adjoint closure, \Cref{app:nn_arch}, with the input layer widened to admit ten features; see below) maps the components of the normalised strain $\mathbf{S}^*$ and rotation $\boldsymbol{\Omega}^*$ tensors, together with $\log(1{+}\mathrm{Re}_T)$, to three momentum body forces $f(\eta)=(f_u, f_v, f_w)$, which enter the RANS momentum equations as
\begin{equation}
\partial_t U_i + U_j \partial_j U_i
 = - \partial_i P
 + \partial_j \left[(\nu+\nu_t)\bigl(\partial_j U_i + \partial_i U_j\bigr)\right]
 + f_i(\eta),
\label{eq:source_term}
\end{equation}
replacing the closure model with a learned body force. The eddy viscosity $\nu_t$ is retained: a pure body-force closure ($\nu_t{=}0$) leaves the momentum equations with no turbulent dissipation and the forward solver does not converge at the Reynolds numbers considered. To supply $\nu_t$, the default Wilcox $k$-$\omega$ transport equations are solved concurrently with fixed coefficients and advance alongside the momentum system; they carry no learnable parameters, but the adjoint still propagates through them, so the body force's indirect effect on $\nu_t$ via $k$-$\omega$ production and dissipation enters the gradient. The final layer of the MLP is scaled by $\epsilon_{\mathrm{init}}{=}10^{-2}$ so that $f_i\approx\mathbf{0}$ and $\nu_t$ reduces to its default-$k$-$\omega$ value at initialisation.

\paragraph{Why raw tensor components rather than invariants.}
Using the same six $(\mathbf{S}, \boldsymbol{\Omega})$-invariants as DARSM fails on the square duct by symmetry. The duct cross-section has dihedral $D_4$ symmetry (four rotations and four reflections); the invariants are scalars under this group, but the body force is a vector with opposite-sign values across the symmetries. A network mapping invariants to body force therefore outputs the average, which is zero. We instead feed the network the nine independent components of $\mathbf{S}^*$ and $\boldsymbol{\Omega}^*$ together with $\log(1+\mathrm{Re}_T)$, which are not invariant under $D_4$ and so let the network produce the correct antisymmetric forcing. DARSM avoids this problem because the Pope-basis tensor structure inside the closure carries the rotational equivariance, so the network outputs only invariant-dependent scalars and the same six invariants suffice as inputs.

\paragraph{Training and deployment.}
The body-force network is trained by the same steady-state adjoint used by our method (\Cref{app:training_procedure}) to minimise $\J_\text{vel}$~\eqref{eq:Jvel}, under the training/validation split of \Cref{sec:comparison}. We checkpoint the best-validation model and report its evaluation on all cases (\Cref{tab:duct}). The same checkpoint is deployed without retraining on the five periodic-hill geometries (\Cref{tab:hills}). The forcing it predicts is not constrained to remain physical outside its training distribution, and its error on the hills is an order of magnitude above the default closure.

\subsection{End-to-end training of the tensor-basis network}
\label{app:tbnn_e2e}

To separate the effect of the parameterisation from that of solver-in-the-loop training, we trained the TBNN of \Cref{app:baseline_apriori} end-to-end on the square duct with the identical objective, optimiser, and adjoint as DARSM; \Cref{sec:comparison} summarises the outcome. The deployment gives the TBNN the same numerical treatment as the EARSM, with the strain-aligned part of the stress treated implicitly as an eddy viscosity and the remainder explicit. The TBNN could not be trained. From a small-output initialisation the network gives $C_\mu\approx0$, which collapses the effective eddy viscosity far below the $k$--$\omega$ value, and the first forward solve diverges at every network width. Initialising the network at the Boussinesq limit, the analogue of DARSM's pretraining to the default constants, allows training to start, but the error remains well above DARSM's and the optimiser soon leaves the region in which the forward solver is stable, after which the solves diverge.

Both closures use the same invariants and tensor basis, so both are frame-invariant, but their networks output different quantities. The TBNN's ten coefficients are unconstrained, whereas DARSM's network outputs only the empirical constants of an equation derived from the Reynolds stress transport equations, so its stress keeps that structure. Consistent with this, \citet{zhang2022ensemble} report that adjoint-based training of the TBNN of \citet{stromer2021end} diverged beyond a linear model.

\subsection{Baselines on the periodic hills}
\label{app:baselines_hills}

Every baseline of \Cref{tab:duct} is also evaluated on the five periodic-hill geometries (\Cref{tab:hills}). Models trained on the square duct (DARSM, DeepONet, FIML's Step-2 network, the source-term closure) are deployed on the hills without retraining. The a-priori NN-EARSM and TBNN are trained offline on the $\alpha{=}1.0$ hills DNS with the protocol of \Cref{app:baseline_apriori}; the NN-EARSM is deployed iteratively and reported at the final iterate where the solver does not converge, and the TBNN, whose iterative deployment diverges, is reported under the frozen-injection protocol. FIML (inversion) is the per-geometry field-inversion step run against each geometry's own DNS and is therefore in sample; its Step-2 network trained on the duct does not reach a steady state out of sample and is reported at the final iterate. The PINN of \Cref{app:baseline_pinn} does not converge on any hill geometry.

\subsection{Scaling benchmark}
\label{app:scaling}

\paragraph{Setup.}
All three routes were evaluated at grids $N\in\{16^2, 32^2, 64^2, 128^2\}$ on CPU in double precision with a $128\,$GB memory ceiling. The neural closure is the production $H{=}10$ gated MLP with six invariant inputs. At each grid, the forward solver is converged once to steady state, and all three routes compute the gradient of $\J$ at the same $\vu$. Peak memory is the increase in resident memory during the gradient computation.

\paragraph{Method definitions.}
\textsc{Unroll} backpropagates through $T$ forward steps (all JIT tridiagonal solvers replaced by autograd-compatible equivalents), bounded by the memory ceiling; we report the maximum depth $T$ reached. \textsc{IFT-AD} solves the implicit adjoint of \Cref{app:adjoint} with the $(\partial\R/\partial\vu)^\top$ matrix-vector product delivered by full reverse-mode AD. \textsc{IFT-Hybrid} uses the production adjoint: structured transpose-Thomas solves plus scalar-contraction autodiff per sub-step. Both \textsc{IFT} routes use GMRES tolerance $10^{-6}$ capped at $2000$ iterations and Poisson BiCGSTAB tolerance $10^{-6}$ capped at $200$ iterations.